\documentclass[12pt]{iopart}
\usepackage{amssymb}
\usepackage{graphicx}
\usepackage{subcaption}

\usepackage[superscript]{cite}
\usepackage[colorlinks=true,linkcolor=blue,urlcolor=blue,citecolor=blue]{hyperref}

\usepackage{longtable}
\usepackage{multirow}

\usepackage{lscape}

\usepackage{setspace}
\begin{document}

\title[Oxygen VUV emission in Ar/O$_2$ ICPs]{Numerical investigation of vacuum ultra-violet emission in Ar/O$_2$ inductively coupled plasmas}

\author{Michel Osca Engelbrecht$^1$, Jonathan Jenderny$^2$, Henrik Hylla$^{2,3}$, Dominik Filla$^{2,3}$, Peter Awakowicz$^2$, Ihor Korolov$^2$, Christopher P. Ridgers$^1$ and Andrew R. Gibson$^3$}

\address{$^1$ York Plasma Institute, School of Physics, Engineering and Technology, University of York, York, UK}
\address{$^2$ Chair of Electrodynamics and Plasma Technology, Faculty of Electrical Engineering and Information Technology, Ruhr University Bochum, Bochum, Germany}
\address{$^3$ Research Group for Biomedical Plasma Technology, Faculty of Electrical Engineering and Information Technology, Ruhr University Bochum, Bochum, Germany}

\ead{michel.osca@york.ac.uk}
\ead{andrew.gibson@rub.de}
\vspace{10pt}
\begin{indented}
\item[]February 2023
\end{indented}

\begin{abstract}
  Controlling ﬂuxes of vacuum ultraviolet (VUV) radiation is important in a number of industrial and biomedical applications of low pressure plasma sources because, depending on the process, VUV radiation may be desired, required to a certain degree, or unwanted. In this work, the emission of VUV radiation from O atoms is investigated in low-pressure Ar/O$_2$ inductively coupled plasmas via numerical simulations. For this purpose, a self-consistent Ar/O$_2$ plasma-chemical reaction scheme has been implemented in a zero dimensional plasma chemical kinetics model and is used to investigate VUV emission from excited O atoms (3s $^5$S$^0_2$ and 3s $^3$S$^0_1$) at 130 and 135\,nm. The model is extensively compared with experimental measurements of absolute VUV emission intensities, electron densities and Ar excited state densities. In addition, VUV emission intensities are investigated as a function of pressure, Ar/O$_2$ mixture, and power deposition and the dominant reaction pathways leading to VUV emission are identiﬁed and described. In general terms, absolute VUV emission intensities increase with power and oxygen fraction over the ranges investigated and peak emission intensities are found for pressures between 5-50\,Pa. The emission is dominated by the 130\,nm resonance line from the decay of the O(3s $^3$S$^0_1$) state to the ground state. Besides, at low pressure (0.3-1\,Pa), the flux of VUV photons to surfaces is much lower than that of positive ions, whereas VUV fluxes dominate at higher pressure, $\gtrsim$5-50\,Pa depending on O$_2$ fraction. Finally, oxygen atom fluxes to surfaces are, in general, larger than those of VUV photons for the parameter space investigated.
\end{abstract}

%\maketitle

\section{Introduction}
\label{sec:introduction}
Inductively coupled plasmas (ICPs) operated at low pressures are widely used for materials processing, microelectronics manufacturing\cite{wang_2001,athavale_1995,schaepkens_2000,barela_2005,banna_2009,agarwal_2011,diomede_2011} and are also investigated for applications in biomedicine\cite{fiebrandt_2018_1, abreu_2013,gomez_lopez_2008, denis_2012, fraise_2011,siani_2015,fiebrandt_2018}.
Control of vacuum ultraviolet (VUV) radiation in ICPs is important as, depending on the process, radiation may be desired\cite{fiebrandt_2018,lerouge_2000} required to some degree\cite{nest_2008,titus_2011,shin_2012} or unwanted\cite{lee_2013, uchida_2008}.
On the one hand, damage to the substrate by VUV radiation during plasma etching can be an important process in materials processing applications and is therefore an active topic of research\cite{tian_2015}.
Otherwise, in some specific circumstances, VUV radiation can participate in synergistic processes\cite{nakano_2002,nest_2008,titus_2011}, where they can be exploited for the benefit of materials processing.
On the other hand, VUV fluxes may be used for the sterilisation of surfaces and are therefore of great interest in medical, pharmaceutical and food industry applications\cite{abreu_2013,gomez_lopez_2008, denis_2012, fraise_2011,siani_2015}.
In this context, VUV radiation for sterilisation purposes is of increasing interest as it can be an effective mechanism on 3-D, heat-sensitive objects and it enables sterilisation in dry environments, with short exposure times and without toxic residues.

VUV emission in ICPs has been investigated for different gas mixtures and under different operation conditions.
Investigations of VUV radiation have been carried out in ICPs operated with different gases, such as Ar\cite{woodworth_2001,jinnai_2010,titus_2009,boffard_2014_1,boffard_2014_2, fiebrandt_2020},  N$_2$\cite{boffard_2014_2, fantz_2016}, O$_2$\cite{boffard_2014_2, fiebrandt_2020}, He\cite{tian_2015}, H$_2$\cite{wunderlich_2021,boffard_2014_2,fantz_2016}, Xe\cite{boffard_2014_2, tian_2015}, Cl$_2$\cite{tian_2017}, Cl$_2$/BCl$_3$\cite{woodworth_1999} and fluorocarbon gases\cite{woodworth_2001, jinnai_2010}, with either experimental or numerical methods in power ranges between 150 and 1100\,W and total pressure ranges between 1 and 100\,mTorr (0.13-13\,Pa).
However, despite the number of investigations carried out, the understanding of the formation pathways of VUV photons in ICP applications remains relatively limited as the operating parameters investigated are comparatively few.
Therefore, a comprehensive investigation of VUV emission in ICPs that describes the pathways leading to emission over a wide range of operating parameters would be useful to better understand and control ICPs for industrial and biomedical applications.

For this reason, an investigation of oxygen atom VUV emission in low pressure Ar/O$_2$ ICPs over the operating parameters of total pressure $p_T$, power $P_{in}$ and oxygen mixture fraction $\chi_{\mathrm{O}_2}$ is carried out in this work.
Oxygen containing plasmas are widely used in industrial applications\cite{cook_1995,collart_1995,korzec_1995,kelly_2000,depla_2009,mitschker_2015,ries_2018} and are of interest for biomedical\cite{moisan_2013,fiebrandt_2018,lerouge_2000} applications. Therefore, providing a detailed understanding of VUV radiation formed from O atoms in Ar/O$_2$ ICPs and the plasma-chemical pathways leading to it could be useful to improve plasma performance in these applications.
%The work presented is a continuation of the collisional radiative model and reaction scheme developed by Fiebrandt \textit{et al}\cite{fiebrandt_2020}.
In this work, the collisional radiative model developed in \citenum{fiebrandt_2020} has been extended and implemented in a zero-dimensional (0D) plasma chemical-kinetics global model (GM) that allows self-consistent simulations.
The GM enables computationally inexpensive simulations and allows detailed study of plasma-chemical and radiative processes and is therefore well suited to the goals of this investigation.

The GM and reaction scheme for Ar/O$_2$ are presented in section \ref{sec:method_numerical_model}.
In parallel, experimental work has been carried out in order to provide a validation of the simulated plasma properties and is described in section \ref{sec:method_exp}.
The numerical GM results are first compared against experimental measurements carried out in this work, and available from previous studies, in \ref{sec:results_part1} to provide confidence in the numerical model and the reaction scheme used.
In this section simulations of electron densities and temperatures, dissociation fractions, argon metasable densities and absolute emission intensities are compared with experimental measurements.
Following comparison with experimental data, a more extensive numerical investigation is carried out over a wide range of operating conditions in section \ref{sec:results_part2}.
In this section variations of the operating parameters of total pressure ($p_T$ = 0.3-100\,Pa), input power ($P_{in}$ = 100-2000\,W) and oxygen fraction ($\chi_{\mathrm{O}_2}$ = 0-0.2) are conducted and oxygen VUV emission and its formation pathways investigated.
The VUV emission is not only described in absolute values but also in comparison with ion and oxygen atom fluxes at the reactor walls to give a broad context on regimes of interest for optimising plasma processes that may be dependent on the fluxes of each different component to surfaces.

%%%%%%%%%%%%%%%%%%%%%%%%%%%%%%%%%%%%%%%%%%%%%%%%%%%%%%%%%%%%%%%%%%%%%%%%%%%%%%%%
%%%%%%%%%%%%%%%%%%%%%%%%%%%%%%%%%%%%%%%%%%%%%%%%%%%%%%%%%%%%%%%%%%%%%%%%%%%%%%%%
%%%%%%%%%%%%%%%%%%%%%%%%%%%%%%%%%%%%%%%%%%%%%%%%%%%%%%%%%%%%%%%%%%%%%%%%%%%%%%%%
\section{Numerical model description}
\label{sec:method_numerical_model}
The numerical method used for this investigation is a 0D plasma-chemical kinetics GM that solves fluid-based mass and energy balance equations for a system of volume $V$ bounded by a surface area $A$.
Under the assumption that mass and energy are relatively homogeneously distributed in space, time variations of species densities and energies are caused by plasma-chemical reactions, interactions with the system boundaries and input power.
This type of model is widely used in the low temperature plasma research community\cite{hurlbatt_2017,lieberman_2005_ch10} as it enables fast simulations of plasmas with complex chemical reaction schemes and can provide robust insights into the scaling of important plasma parameters under variations of external operating conditions\cite{gudmundsson_1999, gudmundsson_2000, gudmundsson_2007}.

For this work, a GM has been designed and developed in the Julia programming language\cite{bezanson_2017}.
The GM models a cylindrical plasma reactor, of length $L$ and radius $R$, to which power $P_{in}$ is coupled inductively.
The experimental reactor is discussed in more detail in section \ref{sec:method_exp}. %where a sketch of the reactor is shown in figure \ref{fig:schematic_dicp}.

\begin{figure}[h]
  \centering
  \includegraphics[width=\linewidth]{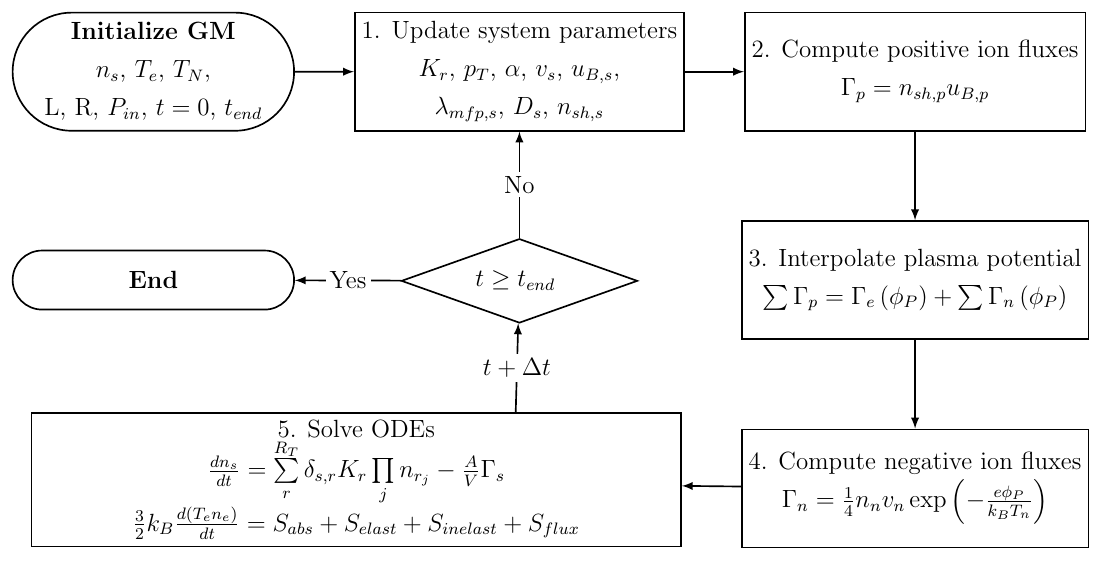}
\caption{\label{fig:GM_flowchart}Flowchart of the 0D plasma-chemical kinetics GM. Steps 2-4 are described in section \ref{sec:method_num_flux_balance} and step 5 in sections \ref{sec:method_num_mass_eq} and \ref{sec:method_num_energy_eq}.}
\end{figure}

The numerical execution structure consists of an initialization of the simulation environment and a five-step cycle, shown in figure \ref{fig:GM_flowchart}, that updates the simulation system in time.

The initial conditions for the density and temperature of each species, as well as the length $L$ and radius $R$ of the reactor, the applied power $P_{in}$ and the simulation time length $t_{end}$ must be defined in advance.
After initialising the simulation parameters, the execution of the cycle computes a new electron temperature and species densities values and advances in time by $\Delta t$.
The simulation ends when the final time condition is reached, $t \ge t_{end}$.
The cycle consists of the following steps
\begin{itemize}
  %\item[0.- ] Set initial conditions: species densities $n_s$, electron temperature $T_e$, neutral gas temperature $T_N$, reactor length $L$ and radius $R$, input power $P_{in}$ and simulation time $t_{end}$.
  \item[1.- ] Computation of system parameters necessary for later steps: reaction rate coefficients $K_r$, species mean-free-path $\lambda_{mfp,s}$, diffusion coefficients $D_s$, electronegativity $\alpha$, total pressure $p_T$, thermal speed $v_s$, Bohm velocity $u_{B,s}$ and number density at the plasma sheath edge $n_{sh,s}$.
  \item[2-4.- ] Computation of charged particle fluxes to the system walls, $\Gamma$, and plasma potential, $\phi_P$. This is described in more detail in section \ref{sec:method_num_flux_balance}.
  \item[5.- ] Solve the system of ordinary differential equations (ODEs) formed by mass and energy continuity equations. The ODE solver \textit{Rosenbrock23} in the \textit{DifferentialEquations} library\cite{rackauckas2017differentialequations} is used for this purpose. A detailed description of the mass and energy equations is found in sections \ref{sec:method_num_mass_eq} and \ref{sec:method_num_energy_eq}, respectively.
\end{itemize}

The simulation results presented in section \ref{sec:results} use the following initial conditions, unless explicitly stated otherwise.
A partially ionised plasma, where the neutral gas is formed only by ground state Ar and O$_2$ at total pressure $p_T$ and with an arbitrary oxygen fraction $0\le \chi_{\mathrm{O}_2}\le 1$.
The plasma is formed by electrons, O$_2^+$ and Ar$^+$ with an initial plasma density $n_P=10^{14}$\,m$^{-3}$.
The initial density of the plasma species fulfils quasi-neutrality, and thus $n_P = n_e = n_{\mathrm{O_2}^+} + n_{Ar^+}$, where $n_{Ar^+} = (1-\chi_{\mathrm{O}_2})n_P$ and $n_{\mathrm{O_2}^+} = \chi_{\mathrm{O}_2}n_P$ are in the same ratio as the Ar/O$_2$ gas mixture.
The remaining neutral and charged species have an initial density of zero.
The initial electron temperature is set to $T_e = 1.5$\,eV and neutral and ion species have a fixed temperature $T_N$. Different values of $T_N$ are used depending on the conditions, as discussed in more detail later.
The reactor size is as the reactor described in the experimental section \ref{sec:method_exp}, with $L=R=$ 0.2\,m.
The simulation time is $t_{end}=$ 1\,s, which has been tested to be long enough for the simulations to converge to a stable solution in all the results presented.

%%%%%%%%%%%%%%%%%%%%%%%%%%%%%%%%%%%%%%%%%%%%%%%%%%%%%%%%%%%%%%%%%%%%%%%%%%%%%%%%
%%%%%%%%%%%%%%%%%%%%%%%%%%%%%%%%%%%%%%%%%%%%%%%%%%%%%%%%%%%%%%%%%%%%%%%%%%%%%%%%
%%%%%%%%%%%%%%%%%%%%%%%%%%%%%%%%%%%%%%%%%%%%%%%%%%%%%%%%%%%%%%%%%%%%%%%%%%%%%%%%
\subsection{Species and plasma-chemical reaction scheme}
\label{sec:chemistry_model}
The species list included in the model is based on previous works on the simulation of argon and oxygen containing plasmas \cite{gudmundsson_2007,turner_2015,fiebrandt_2020}, and is listed in table \ref{tab:species_list}.%: electrons (e), argon (Ar), atomic oxygen (O), molecular oxygen (O$_2$), ozone (O$_3$), tetraoxygen (O$_4$), the corresponding positive and negative ions, and metastable species.
\begin{table}[h]
  \caption{\label{tab:species_list}Species included in the numerical model.}
  \centering
  \begin{tabular}{@{}ll}
  		\br
  		\textbf{Species} & \textbf{Atomic level} \\
  		\mr
      e      &       \\
Ar     &       \\
Ar$^+$ &       \\
Ar(4p) & $\mathrm{3s^2}$ $\mathrm{3p^5(2P^0_{3/2})4p}$, $\mathrm{3s^2}$ $\mathrm{3p^5(2P^0_{1/2})4p}$     \\
Ar$^m$ & $\mathrm{3s^2}$ $\mathrm{3p^5(2P^0_{3/2})4s_2}$, $\mathrm{3s^2}$ $\mathrm{3p^5(2P^0_{1/2})4s_0}$   \\%Ar(1s5) and Ar(1s3)
Ar$^r$ & $\mathrm{3s^2}$ $\mathrm{3p^5(2P^0_{3/2})4s_1}$, $\mathrm{3s^2}$ $\mathrm{3p^5(2P^0_{1/2})4s_1}$  \\%Ar(1s4) and Ar(1s2)
O$_2$   &        \\
O$_2^+$ &        \\
O$_2^-$ &        \\ 
O$_2$(a$^1\Delta_u$)         & \\
%O$_2$(a$^1\Delta_u$,$\nu$)   & \\
O$_2$(b$^1\Sigma^+_u$)       & \\
%O$_2$(b$^1\Sigma^+_u$,$\nu$) & \\
O            & 2s$^2$ 2p$^4$ $^3$P$_{2,1,0}$ \\
O$^+$        & \\
O$^-$        & \\
O($^1$D)        & 2s$^2$ 2p$^4$\ $^1$D$_0$\\
O($^1$S)        & 2s$^2$ 2p$^4$\ $^1$S$_0$\\
O($^3$S)        & 2s$^2$ 2p$^3$ ($^3$S$^0$) 3s $^3$S$^0_1$\\
O($^5$S)        & 2s$^2$ 2p$^3$ ($^3$S$^0$) 3s $^5$S$^0_2$\\
O($^3$P)        & 2s$^2$ 2p$^3$ ($^3$S$^0$) 3p $^3$P$_{1,2,0}$\\
O($^5$P)        & 2s$^2$ 2p$^3$ ($^3$S$^0$) 3p $^5$P$_{1,2,3}$\\
O$_3$        & \\
O$_3$($\nu$) & \\
O$_3^+$      & \\
O$_3^-$      & \\
O$_4$        & \\
O$_4^+$      & \\
O$_4^-$      &  \\
  		\br
  \end{tabular}
\end{table}

The plasma-chemical reaction scheme included in the GM is a compendium of reactions used in \citenum{hurlbatt_2017,fiebrandt_2020,fiebrandt_thesis} and the references therein.
The reaction scheme consists of a set of electron-oxygen reactions, in table \ref{tab:electron_oxygen}, electron-argon, in table \ref{tab:electron_argon}, oxygen-oxygen, in table \ref{tab:oxygen_oxygen}, argon-argon, in table \ref{tab:argon_argon}, oxygen-argon, in table \ref{tab:argon_oxygen}, and recombination reactions, in table \ref{tab:recombination}.
Moreover, additional reactions are included for ion-wall interactions, in table \ref{tab:ion_flux}, neutral-wall interactions, in table \ref{tab:neutral_diffusion}, atomic level transitions, in table \ref{tab:radiative_transitions}, and oxygen reactions with radiative cascading processes, in table \ref{tab:oxygen_radiation}.
Altogether there are a total of $R_T = 393$ reactions included. As is noted in the appendices, reaction rate coefficients for electron impact reactions are implemented as functions of electron temperature, assuming a Maxwellian electron energy distribution function.
%Most of the reactions included are treated as two, or three, body interactions but the reaction scheme also includes particle interactions with the reactor walls and atomic energy level transitions.

Reactions \#13, 31, 50 (in table \ref{tab:electron_oxygen}) and 107 (table \ref{tab:electron_argon}) are electron-neutral elastic collisions.
Reactions \#14, 32 and 51 (table \ref{tab:electron_oxygen}) are rotational excitations, and \#15-20, \#33-38, and \#52-57 (table \ref{tab:electron_oxygen}) are vibrational excitations\cite{turner_2015}, whose products are not explicitly simulated and therefore these reactions only act as an energy gain or loss mechanism.
The reactions \#62 (table \ref{tab:electron_oxygen}) and \#142 (table \ref{tab:electron_argon}) have as product the vibrational state of O$_2$ but this is not included in the model and is replaced by the O$_2$ ground state.

The interactions between electrically charged particles and the reactor walls are described in more detail in section \ref{sec:method_num_flux_balance} and neutral-wall reactions are described in section \ref{sec:method_num_neutral_diffusion}.
Besides, atomic level transitions and radiative processes, especially in oxygen, are described in section \ref{sec:method_num_oxygen_radiation}.

%%%%%%%%%%%%%%%%%%%%%%%%%%%%%%%%%%%%%%%%%%%%%%%%%%%%%%%%%%%%%%%%%%%%%%%%%%%%%%%%
%%%%%%%%%%%%%%%%%%%%%%%%%%%%%%%%%%%%%%%%%%%%%%%%%%%%%%%%%%%%%%%%%%%%%%%%%%%%%%%%
%%%%%%%%%%%%%%%%%%%%%%%%%%%%%%%%%%%%%%%%%%%%%%%%%%%%%%%%%%%%%%%%%%%%%%%%%%%%%%%%
\subsection{Mass balance equations}
\label{sec:method_num_mass_eq}
The basic formulation of the equations used in the model is adapted from Refs. \cite{hurlbatt_2017, gudmundsson_2007}.
The GM includes a mass balance equation for each species $s$, in table \ref{tab:species_list},
\begin{equation}
  \label{eq:mass_balance}
\frac{d n_s}{d t} = \sum_{r}^{R_T} \delta_{s,r}K_r\prod\limits_{j} n_{r_j} - \frac{A}{V}\Gamma_s.
\end{equation}
The left hand side represents the time variations of the density of the $s$-th species, $n_s$. 
The first term on the right hand side accounts for the particle gain, or loss, due to the $R_T$ reactions listed in tables \ref{tab:electron_oxygen}, \ref{tab:electron_argon}, \ref{tab:oxygen_oxygen}, \ref{tab:argon_argon}, \ref{tab:argon_oxygen}, \ref{tab:recombination}, \ref{tab:neutral_diffusion}, \ref{tab:radiative_transitions} and \ref{tab:oxygen_radiation}.
The second term on the right hand side accounts for mass variations caused by particle fluxes of charges particles to the system walls, $\Gamma_s$, that are described in more detail in section \ref{sec:method_num_flux_balance}.
The surface area $A$ and system volume $V$ are determined by the cylindrical shape of the reactor, i.e. $A = 2\pi (R^2 + RL)$ and $V = \pi R^2L$.

The mass variation caused by the $r$-th reaction is the product of the rate coefficient $K_r$ with the densities of the $j$ reacting species, $n_{r_j}$.
The factor $\delta_{s,r}$ is an integer that reflects the particle balance of species $s$ in reaction $r$.
For instance, in reaction \#1 ($e + \mathrm{O}\to 2e + \mathrm{O}^+$ in table \ref{tab:electron_oxygen}) electrons have a positive balance $\delta_{1,e} = 1$, atomic oxygen a negative balance $\delta_{1,\mathrm{O}}=-1$, and oxygen ions a positive balance $\delta_{1,\mathrm{O}^+}=1$.
Essentially, $\delta_{s,r} < 0$ represents a mass loss, $\delta_{s,r} > 0$ gain, and $\delta_{s,r} = 0$ equilibrium.

%%%%%%%%%%%%%%%%%%%%%%%%%%%%%%%%%%%%%%%%%%%%%%%%%%%%%%%%%%%%%%%%%%%%%%%%%%%%%%%%
%%%%%%%%%%%%%%%%%%%%%%%%%%%%%%%%%%%%%%%%%%%%%%%%%%%%%%%%%%%%%%%%%%%%%%%%%%%%%%%%
%%%%%%%%%%%%%%%%%%%%%%%%%%%%%%%%%%%%%%%%%%%%%%%%%%%%%%%%%%%%%%%%%%%%%%%%%%%%%%%%
\subsection{Energy conservation equation}
\label{sec:method_num_energy_eq}
The energy balance equation accounts for changes in species temperatures as a function of time.
The energy balance equation is only solved for electrons, while the temperatures of heavy particles are assumed to be constant in time.
%The ion and neutral terms in the energy equation can be neglected due to its larger mass.
%This means that the low inertia of electrons allows a fast response to energy variations and therefore electrons are able to reach a local thermal equilibrium much faster than, and independent of, ions and neutrals\cite{griem_1997}.
%Species with higher mass have an impact on the overall energy balance, however, this is reached on time scales much longer than the chemical kinetics of interest and can therefore be neglected and assumed that their EDF are static.
Here, the shape of the EDF of electrons is assumed to be Maxwellian. The potential limitations of this assumption are discussed further later. The energy equation for electrons takes the following form
%Under these assumptions, ions and neutrals have a constant temperature, and only electrons are included in the energy balance equation
\begin{equation}
\frac{3}{2}k_B\frac{d (T_en_e)}{d t} = S_{abs} + S_{elast} + S_{inelast} + S_{flux},
\label{eq:energy_balance}
\end{equation}
where the electron temperature $T_e$ is used as energy reference parameter, $k_B$ is the Boltzmann constant, $n_e$ is the electron density, $S_{abs}$ is the input power absorbed per unit volume, $S_{elast}$ represents energy changes caused by elastic collision processes, $S_{inelast}$ are energy changes caused by inelastic and superelastic collision processes, and $S_{flux}$ is related to the kinetic energy lost by electron and ion fluxes through the plasma sheath.

The input power absorption rate in equation \ref{eq:energy_balance}
\begin{equation}
S_{abs} = \frac{P_{in}}{V},
\end{equation}
represents the external inductive power $P_{in}$ that is coupled to the electrons.

The term $S_{elast}$ represents the electron energy gains and losses caused by elastic collisions, of the type $e + N \to e + N$ where $N$ is a neutral species,
\begin{equation}
S_{elastic} = -3 \sum_{l}^{R_{elast}} \frac{m_e}{m_N}k_B\left(T_e - T_N\right)K_{l}n_e n_N,
\end{equation}
where $R_{elast}$ is the set of elastic collisions present in the collision model, $m_N$ is the mass of $N$, $T_N$ is the temperature of $N$, and $K_l$ is the rate coefficient of the $l$-th elastic scattering process.

Gains or losses of energy caused by inelastic and superelastic collision processes are accounted as
\begin{equation}
S_{inelast} = -\sum_r E_{thr,r}K_r\prod\limits_{r_j} n_{j}.
\end{equation}
where $E_{thr,r}$ is the energy released, or absorbed, by the $r$-th collision.
%This term is applied to the reactions listed in tables \ref{tab:electron_oxygen} to \ref{tab:oxygen_radiation}.

%Gains or losses of energy caused by production of consumption of electrons are quantified similarly to equation \ref{eq:mass_balance},
%\begin{equation}
%S_{mass} = -\frac{3}{2} k_B T_e \sum_r^{R_T}\delta_{e,r}K_r\prod\limits_{r_j} n_{j}.
%\end{equation}

The last term in equation \ref{eq:energy_balance} accounts for the kinetic energy of electrons and positive ions that pass through the sheath and are lost at surfaces
\begin{equation}
S_{Flux} = -\frac{A}{V}\left[2k_B T_e\Gamma_e + \sum_{p}\Gamma_{p}\left(\frac{1}{2}k_B T_e + q_{p}\phi_P\right)\right],
\label{eq:energy_term_flux}
\end{equation}
where $\Gamma$ is the particle flux at the system walls, the subscript $p$ is for positive ions, $\phi_P$ is the plasma potential, and $q_p$ is electric charge.
The first term on the right hand side accounts for the kinetic energy taken to surfaces by electrons that have passed through the sheath and the second term accounts for the kinetic energy taken to surfaces by positive ions that have passed across the sheath\cite{lieberman_2005_ch6}.
How particle fluxes crossing the sheath are handled in the GM is described in more detail in the following section.

%%%%%%%%%%%%%%%%%%%%%%%%%%%%%%%%%%%%%%%%%%%%%%%%%%%%%%%%%%%%%%%%%%%%%%%%%%%%%%%%
%%%%%%%%%%%%%%%%%%%%%%%%%%%%%%%%%%%%%%%%%%%%%%%%%%%%%%%%%%%%%%%%%%%%%%%%%%%%%%%%
%%%%%%%%%%%%%%%%%%%%%%%%%%%%%%%%%%%%%%%%%%%%%%%%%%%%%%%%%%%%%%%%%%%%%%%%%%%%%%%%
\subsection{Ion fluxes to the reactor walls}
\label{sec:method_num_flux_balance}
Ion fluxes crossing the plasma sheaths and reaching the reactor walls play an important role in the mass and energy balance equations.
Moreover, ion fluxes are also important to compute the plasma potential $\phi_P$, which is required for the electron energy equation and for fluxes of negatively charged species.
Positive ion (subscript $p$) fluxes are computed differently from negative ion (subscript $n$) and electron (subscript $e$) fluxes.

Positive ions, whose fluxes are given by
\begin{equation}
\Gamma_p = n_{sh,p} u_{B,p},
\label{eq:flux_ions_positive}
\end{equation}
where $n_{sp,p}$ is the density at the sheath, need to enter the sheath with the Bohm velocity $u_{B,p} = \sqrt{k_BT_e/m_p}$ in order to be able to reach the walls.
The effective density at the sheath edge\cite{gudmundsson_1999, lieberman_2005_ch5}
\begin{equation}
  n_{sh,p} = \frac{R^2h_{L,p} + RLh_{R,p}}{R^2 + RL}n_p
  \label{eq:dens_presheath}
\end{equation}
is determined from bulk plasma densities, $n_p$ using geometrical factors $R$ and $L$ as well as the parameters\cite{thorsteinsson_2010}
\begin{equation}
    h_{\{R,L\},p} = \left[\left(\frac{h_{\{R,L\}0}}{1+3\alpha/2}\right)^2 + h_c^2\right]^{1/2}
\end{equation}
where
\begin{equation}
h_{R0,p} = 0.8 \left[4+\frac{\eta R}{ \lambda_{mfp,p}} + \left(\frac{0.8Ru_{B,p}}{\chi_{01}J_1(\chi_{01})D_{a,p}}\right)^2\right]^{-1/2},
\end{equation}
\begin{equation}
h_{L0,p} = 0.86\left[3+\frac{\eta L}{2\lambda_{mfp,p}} + \left(\frac{0.86Lu_{B,p}}{\pi D_{a,p}} \right)^2 \right]^{-1/2},
\end{equation}
\begin{equation}
    h_c = \frac{1}{\gamma_-^{1/2} + \gamma_+^{1/2}[n_{*,p}^{1/2}n_+/n_{-}^{3/2}]}.
\end{equation}
These parameters enable the computation of the sheath edge density from very low pressure regimes, where the ion mean free path is much larger than the system dimensions $\lambda_{mfp,p}\gg (L,R)$, to high pressures, where $\lambda_{mfp}\ll T_e/T_p(R,L)$\cite{lieberman_2005_ch5, lee_1995}.
The $h_{\{R,L\}0}$ parameters make use of $\chi_{01}\simeq 2.405$, the first zero of the zero order Bessel function $J_0$, and the Bessel function 1 of the first kind $J_1$. The plasma electronegativity is given by
\begin{equation}
\alpha = \frac{1}{n_e}\sum_n n_{n}.
\end{equation}
The temperature ratio between positive and negative ions is given by
\begin{equation}
    \eta = \frac{2T_+}{T_+ + T_-},
\end{equation}
where the subscript $+$ and $-$ refer to all positive and negative ion species, respectively. The ambipolar diffusion coefficient is calculated as
\begin{equation}
    D_{a,p} = D_p\frac{1 + \gamma_p + \gamma_p\alpha}{1 + \gamma_p\alpha}
\end{equation}
where \begin{equation}
  \gamma_p = T_e / T_p,
\end{equation}
is the temperature ratio between electrons and the ion species. 

The diffusion coefficient for ions (and also for neutrals, as discussed in the next section) is defined as
\begin{equation}
  D_p = \frac{1}{\sum\limits_s\frac{1}{D_{ps}}} % = \frac{1}{\sum\limits_r\frac{k_BT}{\mu\nu_r}}
  \label{eq:Dp}
\end{equation}
which represents an approximation for the diffusion of a species in a multicomponent mixture. Here, $D_{ps} = k_B T_N / \mu_{ps} \nu_{ps}$ is the binary diffusion coefficient\cite{lieberman_2005_ch9} between the given ion $p$ and the $s$-th heavy mass species in the system, i.e. species with $m_s\gg m_e$. %Besides, $\bar{v_{pi}} = \sqrt{8k_BT_{p}/\pi\mu_{ps}}$ is the mean speed of relative energy\cite{lieberman_2005_ch9}
 Besides, $\nu_{ps} = n_s\sum\limits_{r} K_r$ is the total collision frequency between $p$ and $s$,  and $\mu_{ps}$  is the reduced mass. 

The $h_c$ parameter makes use of $\gamma_- = T_e / T_-$ and $\gamma_+ = T_e / T_+$, which in our case are the same as the temperature of ions and neutrals are equal $T_- = T_+ = T_N$,  and
\begin{equation}
    n_{*,p} = \frac{15}{56}\frac{\eta^2}{K_{rec}\lambda_{mfp,p}}v_p,
\end{equation}
where $K_{rec}$ is the total rate coefficient of the recombination reactions listed in table \ref{tab:recombination}.

The total mean-free-path is estimated as
\begin{equation}
\lambda_{mfp, p} =   \frac{1}{\sum\limits_s n_s\sigma^T_{ps} }
\label{eq:mfp}
\end{equation}
where $\lambda_{mfp,ps} = 1/n_s\sigma^T_{ps}$ and $\sigma^T_{ps}$ is the total collision cross-section between species $p$ and $s$.
Please note that $s$ refers only to heavy mass species, and therefore the corresponding neutral-ion and ion-ion collisions listed in tables \ref{tab:oxygen_oxygen}-\ref{tab:recombination}, as well as elastic scattering, resonant charge-exchange and Coulomb collision processes are included in the calculation of the mean-free-path.
The cross-section of the reactions in the above-mentioned tables are approximated with  $\sigma_{ps} \simeq K_{r}/v_{ps}$\cite{lieberman_2005_ch3} where $v_{ps} = \sqrt{8 k_B T_N /\pi \mu_{ps}}$ is the mean speed of relative motion\cite{lieberman_2005_ch9}.
The cross-section of elastic scattering and resonant charge-exchange are extracted from \citenum{smirnov_1977, lieberman_2005_ch9, gudmundsson_2007}, if available, otherwise they are calculated using the hard sphere model, $\sigma_{ps} = \pi(r_p + r_s)^2$, using the following atomic, and molecular, radii: $r_{\mathrm{Ar}} = 188$\,pm, $r_{\mathrm{O}} = 152$\,pm, $r_{\mathrm{O}_2} = r_{\mathrm{O}_3} = r_{\mathrm{O}_4} = 197$\,pm. For Coulomb collisions, a constant cross-section estimate of $5\cdot10^{-19}$\,m$^{2}$ is used\cite{thorsteinsson_2010}. 

Negative ion fluxes to surfaces are described by the expression given in \citenum{gudmundsson_2000}
\begin{equation}
\Gamma_n = \frac{1}{4}n_n v_{n}\exp{\left(-\frac{e\phi_P}{k_B T_n}\right)},
\label{eq:flux_ions_negative}
\end{equation}
where the subscript $n$ refers to negative ion species.
The flux of these species are restricted to those particles with energies high enough to overcome the potential barrier of the plasma sheath, which is determined in ICPs by the plasma potential with respect to a floating wall.
Note that $v_{n} = \sqrt{8k_BT_n/\pi m_n}$ is the thermal speed of the $n$-th negative ion.
The same expression as in equation \ref{eq:flux_ions_negative} is valid for the electron flux, $\Gamma_e$.
To determine $\Gamma_n$ and $\Gamma_e$ the plasma potential $\phi_P$ must be known, which is obtained by solving the flux balance equation
\begin{equation}
\sum_p q_p\Gamma_p + q_e\Gamma_e + \sum_n q_n\Gamma_n = 0,
\label{eq:flux_balance}
\end{equation}
which states that the total particle flux, of positive, negative ions and electrons, must balance to ensure quasi-neutrality.
The flux balance equation is solved for $\phi_P$ using an iterative method.
$\phi_P$ is then used in the flux term of the energy balance equation, equation \ref{eq:energy_term_flux}, and for computing the flux of negative ions and electrons, equation \ref{eq:flux_ions_negative}.

In order to maintain mass conservation in the system, both positive and negative ions are considered to be neutralised when they get in contact with the wall\cite{gudmundsson_2007}.
These reactions are listen in table \ref{tab:ion_flux}, such that $A/V\Gamma_s = \delta_{s,r}n_sK_r$\cite{gudmundsson_2007}, and are included in the mass balance (second term on rhs of equation \ref{eq:mass_balance}) for the species on both left and right sides of the neutralization reactions.
Note that the ion-wall neutralization reactions in Ref. \citenum{gudmundsson_2007} have been extended to the ion species included in this work.
\begin{table}[h]
  \caption{\label{tab:ion_flux}Ion-wall reactions.}
  \centering
	\begin{tabular}{@{}llll}
 		\br
 		\textbf{\#} & \textbf{Process} & \textbf{$K_r$ [s$^{-1}$]} & \textbf{Ref.} \\
 		\mr
     344 & $\mathrm{O}^+ \to \mathrm{O}$ & $2u_{B,\mathrm{O}^+} \big(R^2 h_{L,\mathrm{O}^+} + R L h_{R,\mathrm{O}^+} \big) / \big(R^2 L\big) $ & \citenum{gudmundsson_2007} \cr
345 & $\mathrm{O}_2^+ \to \mathrm{O}_2$ & $2u_{B,\mathrm{O}_2^+}\big(R^2 h_{L,\mathrm{O}_2^+} + R L h_{R,\mathrm{O}_2^+}\big) / \big(R^2 L\big) $ & \citenum{gudmundsson_2007} \cr
346 & $\mathrm{O}_3^+ \to \mathrm{O}_3$ & $2u_{B,\mathrm{O}_3^+}\big(R^2 h_{L,\mathrm{O}_3^+} + R L h_{R,\mathrm{O}_3^+}\big) / \big(R^2 L\big) $ & \citenum{gudmundsson_2007}$^a$ \cr
347 & $\mathrm{O}_4^+ \to 2\mathrm{O}_2$ & $2u_{B,\mathrm{O}_4^+}\big(R^2 h_{L,\mathrm{O}_4^+} + R L h_{R,\mathrm{O}_4^+}\big) / \big(R^2 L\big) $ & \citenum{gudmundsson_2007}$^a$ \cr
348 & $\mathrm{Ar}^+ \to \mathrm{Ar}$ & $2u_{B,\mathrm{Ar}^+} \big(R^2 h_{L,\mathrm{Ar}^+} + R L h_{R,\mathrm{Ar}^+} \big) / \big(R^2 L\big) $ & \citenum{gudmundsson_2007} \cr
349 & $\mathrm{O}^- \to \mathrm{O}$ & $(A/4V)v_{\mathrm{O}^{-}} \exp{\big(-e\phi_P / k_BT_{ \mathrm{O}^{-}}\big)}$ & \citenum{gudmundsson_2000}\cr
350 & $\mathrm{O}_2^- \to \mathrm{O}_2$ & $(A/4V)v_{\mathrm{O}_2^{-}}\exp{\big(-e\phi_P / k_BT_{\mathrm{O}_2^{-}}\big)}$ & \citenum{gudmundsson_2000}$^b$ \cr
351 & $\mathrm{O}_3^- \to \mathrm{O}_3$ & $(A/4V)v_{\mathrm{O}_3^{-}}\exp{\big(-e\phi_P / k_BT_{\mathrm{O}_3^{-}}\big)}$ & \citenum{gudmundsson_2000}$^b$ \cr
352 & $\mathrm{O}_4^- \to 2\mathrm{O}_2$ & $(A/4V)v_{\mathrm{O}_4^{-}}\exp{\big(-e\phi_P / k_BT_{\mathrm{O}_4^{-}}\big)}$ & \citenum{gudmundsson_2000}$^b$

     \\
 		\br
    \multicolumn{4}{l}{\footnotesize$^a$ The expression is of the same form given in Ref. \citenum{gudmundsson_2007}, but is extended here to all positively charged species} \\\multicolumn{4}{l}{\footnotesize$^b$ The expression is of the same form given in Ref. \citenum{gudmundsson_2000}, but is extended here to all negatively charged species} 
	\end{tabular}
\end{table}

%%%%%%%%%%%%%%%%%%%%%%%%%%%%%%%%%%%%%%%%%%%%%%%%%%%%%%%%%%%%%%%%%%%%%%%%%%%%%%%
%%%%%%%%%%%%%%%%%%%%%%%%%%%%%%%%%%%%%%%%%%%%%%%%%%%%%%%%%%%%%%%%%%%%%%%%%%%%%%%
%%%%%%%%%%%%%%%%%%%%%%%%%%%%%%%%%%%%%%%%%%%%%%%%%%%%%%%%%%%%%%%%%%%%%%%%%%%%%%%
\subsection{Neutral particle diffusion to the reactor walls}
\label{sec:method_num_neutral_diffusion}
Neutral particle diffusion within the plasma reactor plays an important role as it determines the flux of neutral species that interact with the reactor walls\cite{gudmundsson_1999, lieberman_2005_ch5,lieberman_2005_ch10}.
This is important because metastable species reaching the walls are de-excited to ground state, and atomic oxygen recombines into molecular oxygen.
Therefore, neutral-wall interactions depend on the species diffusion properties.
These types of reactions are included in the GM, and listed in table \ref{tab:neutral_diffusion}.
\begin{table}[h]
  \caption{\label{tab:neutral_diffusion}Neutral-wall reactions. $\gamma$ is the sticking coefficient.}
  \centering
    \begin{tabular}{@{}lllll}
		 \br
		 \textbf{\#} & \textbf{Process} & \textbf{$\gamma$} &\textbf{$K_r$ [s$^{-1}$]} & \textbf{Ref.} \\
		 \mr
 	%  # Wall reactions (Gudmundsson 2007) 
353 & $\mathrm{O} \to \frac{1}{2}\mathrm{O}_2$ & equation \ref{eq:sticking_O} & $\bigg[\frac{\Lambda^2}{D_\mathrm{O}}+ \frac{2V(2 - \gamma_{\mathrm{O}} )}{A v_{\mathrm{O}} \gamma_{\mathrm{O} }}\bigg]^{-1}$ & \citenum{gudmundsson_2007} \cr
354 & $\mathrm{O(^1D)} \to \mathrm{O}$ & 1.0 & $\bigg[\frac{\Lambda^2}{D_{\mathrm{O(^1D)}}} + \frac{2V(2 - \gamma_{\mathrm{O(^1D)}} )}{A v_{\mathrm{O(^1D)}} \gamma_{\mathrm{O(^1D)} }}\bigg]^{-1}$ & \citenum{fiebrandt_2020} \cr
355 & $\mathrm{O(^1S)} \to \mathrm{O}$ & 1.0 & $\bigg[\frac{\Lambda^2}{D_{\mathrm{O(^1S)}}} + \frac{2V(2 - \gamma_{\mathrm{O(^1S)}} )}{A v_{\mathrm{O(^1S)}} \gamma_{\mathrm{O(^1S)} }}\bigg]^{-1}$ & \citenum{fiebrandt_2020} \cr
356 & $\mathrm{O(^3S)} \to \mathrm{O}$ & 1.0 & $\bigg[\frac{\Lambda^2}{D_{\mathrm{O(^3S)}}} + \frac{2V(2 - \gamma_{\mathrm{O(^3S)}} )}{A v_{\mathrm{O(^3S)}} \gamma_{\mathrm{O(^3S)} }}\bigg]^{-1}$ & \citenum{fiebrandt_2020} \cr
357 & $\mathrm{O(^5S)} \to \mathrm{O}$ & 1.0 & $\bigg[\frac{\Lambda^2}{D_{\mathrm{O(^5S)}}} + \frac{2V(2 - \gamma_{\mathrm{O(^5S)}} )}{A v_{\mathrm{O(^5S)}} \gamma_{\mathrm{O(^5S)} }}\bigg]^{-1}$ & \citenum{fiebrandt_2020} \cr
358 & $\mathrm{O(^3P)} \to \mathrm{O}$ & 1.0 & $\bigg[\frac{\Lambda^2}{D_{\mathrm{O(^3P)}}} + \frac{2V(2 - \gamma_{\mathrm{O(^3P)}} )}{A v_{\mathrm{O(^3P)}} \gamma_{\mathrm{O(^3P)} }}\bigg]^{-1}$ & \citenum{fiebrandt_2020} \cr
359 & $\mathrm{O(^5P)} \to \mathrm{O}$ & 1.0 & $\bigg[\frac{\Lambda^2}{D_{\mathrm{O(^5P)}}} + \frac{2V(2 - \gamma_{\mathrm{O(^5P)}} )}{A v_{\mathrm{O(^5P)}} \gamma_{\mathrm{O(^5P)} }}\bigg]^{-1}$ & \citenum{fiebrandt_2020} \cr
360 & $\mathrm{O}_2(\mathrm{a}^1\Delta_u) \to \mathrm{O}_2$ & 0.007 & $\bigg[\frac{\Lambda^2}{D_{\mathrm{O}_2(\mathrm{a}^1\Delta_u)}} + \frac{2V(2 - \gamma_{\mathrm{O}_2(\mathrm{a}^1\Delta_u)} )}{A v_{\mathrm{O}_2(\mathrm{a}^1\Delta_u) } \gamma_{\mathrm{O}_2(\mathrm{a}^1\Delta_u) }}\bigg]^{-1}$ & \citenum{gudmundsson_2007,sharpless_1989} \cr
361 & $\mathrm{O}_2(\mathrm{b}^1\Sigma^+_u) \to \mathrm{O}_2$ & 0.007 & $\bigg[\frac{\Lambda^2}{D_{\mathrm{O}_2(\mathrm{b}^1\Sigma^+_u)}} + \frac{2V(2 - \gamma_{\mathrm{O}_2(\mathrm{b}^1\Sigma^+_u)} )}{A v_{\mathrm{O}_2(\mathrm{b}^1\Sigma^+_u) } \gamma_{\mathrm{O}_2(\mathrm{b}^1\Sigma^+_u) }}\bigg]^{-1}$ & \citenum{gudmundsson_2007,sharpless_1989} \cr
362 & $\mathrm{Ar}^m \to \mathrm{Ar}$ & 1.0 & $\bigg[\frac{\Lambda^2}{D_{\mathrm{Ar}^m}} + \frac{2V(2 - \gamma_{\mathrm{Ar}^m)} )}{A v_{\mathrm{Ar}^m} \gamma_{\mathrm{Ar}^m) }}\bigg]^{-1}$ & \citenum{gudmundsson_2007} \cr
363 & $\mathrm{Ar}^r \to \mathrm{Ar}$ & 1.0 & $\bigg[\frac{\Lambda^2}{D_{\mathrm{Ar}^r}} + \frac{2V(2 - \gamma_{\mathrm{Ar}^r)} )}{A v_{\mathrm{Ar}^r} \gamma_{\mathrm{Ar}^r) }}\bigg]^{-1}$ & \citenum{gudmundsson_2007} \cr
364 & $\mathrm{Ar(4p)} \to \mathrm{Ar}$ & 1.0 & $\bigg[\frac{\Lambda^2}{D_{\mathrm{Ar(4p)}}} + \frac{2V(2 - \gamma_{\mathrm{Ar(4p)}} )}{A v_{\mathrm{Ar(4p)}} \gamma_{\mathrm{Ar(4p)}) }}\bigg]^{-1}$ & \citenum{gudmundsson_2007}

 	\\
		 \br
    \end{tabular}
\end{table}

The effective loss-rate coefficient for a neutral species $N$ to the wall is given by\cite{chantry_1987, booth_1991}
\begin{equation}
  K_{D,N} = \bigg[\frac{\Lambda^2}{D_N}+ \frac{2V(2-\gamma_N)}{Av_N\gamma_N} \bigg]^{-1}
\end{equation}
where
\begin{equation}
  \Lambda = \bigg[(\frac{\pi}{L})^2 + (\frac{2.405}{R})^2\bigg]^{-1/2}
\end{equation}
is the effective diffusion length for a cylindrical reactor\cite{chantry_1987}, $D_N$ is the diffusion coefficient for neutrals, $v_N=\sqrt{8k_BT_N/\pi m_N}$ is the thermal speed and $\gamma_N$ is the sticking coefficient.
$D_N$ and the mean free path $\lambda_{mfp,N}$ are defined as in equations \ref{eq:Dp} and  \ref{eq:mfp} respectively, but for neutrals instead of ions. %, with the neutral mean free path, $\lambda_{mfp,N}$, computed using equation \ref{eq:mfp}.
%In this case, the calculation of $\lambda_{mfp,N}$ includes neutral-neutral elastic collisions.
The sticking coefficient depends, among other parameters, on the wall material and operating pressure\cite{gudmundsson_1999,gudmundsson_2007}.
The GM uses $\gamma_N$ values taken from \citenum{fiebrandt_2020, gudmundsson_2007} that conducted simulations under similar operating conditions.
The $\gamma_N$ values used, listed in table \ref{tab:neutral_diffusion}, are constant parameters except for atomic oxygen\cite{gudmundsson_2007}, which is pressure dependent based on the following expression
\begin{equation}
  \label{eq:sticking_O}
  \gamma_\mathrm{O} = \cases{
    (1 - p_{\mathrm{O}_2}[\mathrm{mTorr}])/4, & $p_{\mathrm{O}_2} < 2$ mTorr \\
    0.1438\exp{(2.5069/p_{\mathrm{O}_2}[\mathrm{mTorr}])}, & otherwise.
  }
\end{equation}

%%%%%%%%%%%%%%%%%%%%%%%%%%%%%%%%%%%%%%%%%%%%%%%%%%%%%%%%%%%%%%%%%%%%%%%%%%%%%%%
%%%%%%%%%%%%%%%%%%%%%%%%%%%%%%%%%%%%%%%%%%%%%%%%%%%%%%%%%%%%%%%%%%%%%%%%%%%%%%%
%%%%%%%%%%%%%%%%%%%%%%%%%%%%%%%%%%%%%%%%%%%%%%%%%%%%%%%%%%%%%%%%%%%%%%%%%%%%%%%
\subsection{Atomic energy transitions and radiative processes}
\label{sec:method_num_oxygen_radiation}

\begin{figure}[h]
  \centering
  \includegraphics[width=\linewidth]{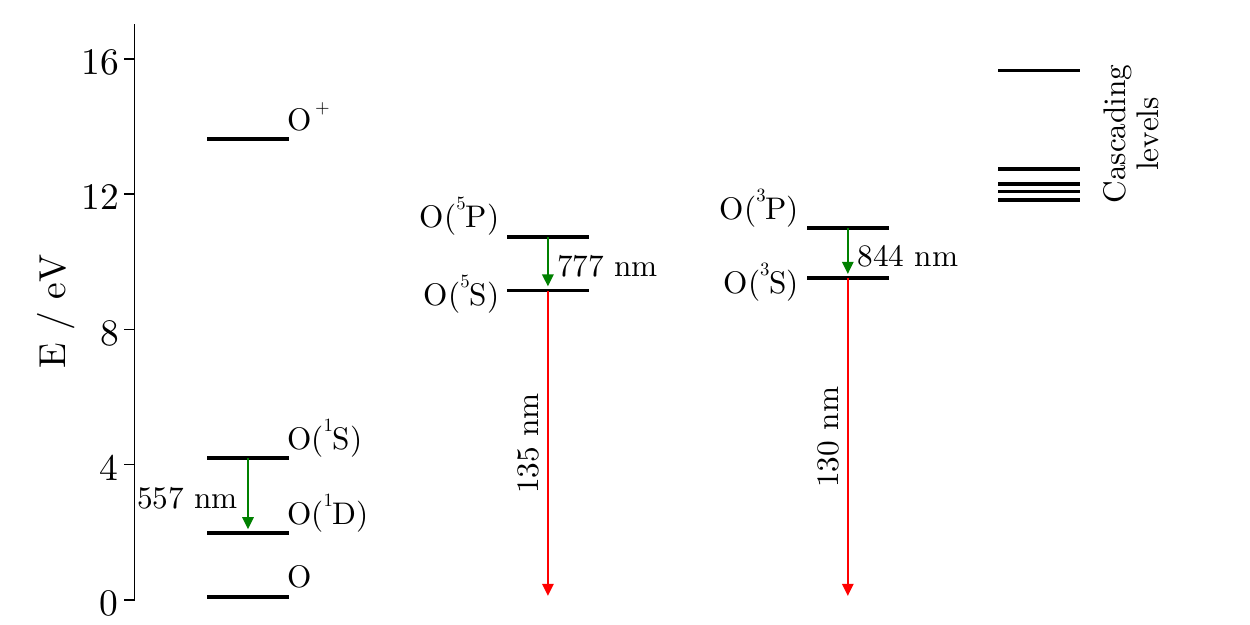}
\caption{\label{fig:O_transitions}Energy diagram of atomic oxygen and radiative transitions taken into account in the numerical model. The cascading levels shown are only a representative subset of the existing high energy levels\cite{kramida_2015}. Figure adapted from Ref. \citenum{fiebrandt_2020}.}
\end{figure}
Radiation processes from certain excited states when they decay to lower energy levels are included in the GM.
The natural decay of excited species at energy level $a$ to a lower energy level $b$ emitting radiation at a wavelength $\lambda_{ab}$ is a well-known physical phenomenon whose rates are described by Einstein coefficients for spontaneous emission.
The radiative reactions included in the GM are sketched in figure \ref{fig:O_transitions} and listed in table \ref{tab:radiative_transitions}.
\begin{table}[h]
  \caption{\label{tab:radiative_transitions}Atomic transitions from state $a \to b$. $\lambda_{ab}$ is the radiation wavelength, $A_{ab}$ is the Einstein coefficient for spontaneous emission, $g_a$ and $g_b$ are the statistical weights of the $a$ and $b$ levels, respectively, and $\gamma_{ab}$ is the escape factor.}
  \centering
    \begin{tabular}{@{}llllllll}
		 \br
		 \textbf{\#} & \textbf{Process} & \textbf{$K_r$ [s$^{-1}$]} & \textbf{$\lambda_{ab}$ [nm]} & \textbf{$A_{ab}$ [s$^{-1}$]} & \textbf{$g_p$} & \textbf{$g_k$} & \textbf{Ref.} \\
		 \mr
	%  # Fiebrandt (2020) table 2 and/or Fiebrandt thesis (2018)
%  # reaction;   [g_p, g_k, g_p_tot, g_k_tot, wavelength];      A_pk;
%  # Ar reactions (Gudmundsson 2007, table 1)
365 & $\mathrm{O(^1S)} \to \mathrm{O(^1D)}$ & $\gamma_{ab} A_{ab}$ & 557.7 & 1.26 & 1.0 & 5.0 & \citenum{fiebrandt_2020,kramida_2015} \cr
366 & $\mathrm{O(^5S)} \to \mathrm{O} $ & $0.5\gamma_{ab} A_{ab}$ & 135.6 & 4.2$\cdot 10^3$ & 5.0 & 5.0 & \citenum{fiebrandt_2020,kramida_2015} \cr
367 & $\mathrm{O(^5S)} \to \mathrm{O} $ & $0.5\gamma_{ab} A_{ab}$ & 135.9 & 1.4$\cdot 10^3$ & 5.0 & 3.0 & \citenum{fiebrandt_2020,kramida_2015} \cr
368 & $\mathrm{O(^3S)} \to \mathrm{O} $ & $0.33\gamma_{ab} A_{ab}$ & 130.2 & 3.4$\cdot 10^8$ & 3.0 & 5.0 & \citenum{fiebrandt_2020,kramida_2015} \cr
369 & $\mathrm{O(^3S)} \to \mathrm{O} $ & $0.33\gamma_{ab} A_{ab}$ & 130.5 & 2.0$\cdot 10^8$ & 3.0 & 3.0 & \citenum{fiebrandt_2020,kramida_2015} \cr
370 & $\mathrm{O(^3S)} \to \mathrm{O} $ & $0.33\gamma_{ab} A_{ab}$ & 130.6 & 6.8$\cdot 10^7$ & 3.0 & 1.0 & \citenum{fiebrandt_2020,kramida_2015} \cr
371 & $\mathrm{O(^5P)} \to \mathrm{O(^5S)}$ & $0.47\gamma_{ab} A_{ab}$ & 777.2 & 3.7$\cdot 10^7$ & 7.0 & 5.0 & \citenum{fiebrandt_2020,kramida_2015} \cr
372 & $\mathrm{O(^5P)} \to \mathrm{O(^5S)}$ & $0.33\gamma_{ab} A_{ab}$ & 777.4 & 3.7$\cdot 10^7$ & 5.0 & 5.0 & \citenum{fiebrandt_2020,kramida_2015} \cr
373 & $\mathrm{O(^5P)} \to \mathrm{O(^5S)}$ & $0.2\gamma_{ab} A_{ab}$ & 777.5 & 3.7$\cdot 10^7$ & 3.0 & 5.0 & \citenum{fiebrandt_2020,kramida_2015} \cr
374 & $\mathrm{O(^3P)} \to \mathrm{O(^3S)}$ & $0.11\gamma_{ab} A_{ab}$ & 844.6 & 9.2$\cdot 10^7$ & 1.0 & 3.0 & \citenum{fiebrandt_2020,kramida_2015} \cr
375 & $\mathrm{O(^3P)} \to \mathrm{O(^3S)}$ & $0.56\gamma_{ab} A_{ab}$ & 844.6 & 9.2$\cdot 10^7$ & 5.0 & 3.0 & \citenum{fiebrandt_2020,kramida_2015} \cr
376 & $\mathrm{O(^3P)} \to \mathrm{O(^3S)}$ & $0.33\gamma_{ab} A_{ab}$ & 844.7 & 9.2$\cdot 10^7$ & 3.0 & 3.0 & \citenum{fiebrandt_2020,kramida_2015} \cr
377 & $\mathrm{Ar}^r \to \mathrm{Ar}$ & $A_{ab}$ & & $10^5$ & & & \citenum{gudmundsson_2007, hurst_2003} \cr
378 & $\mathrm{Ar(4p)} \to \mathrm{Ar}$ & $A_{ab}$ & & 3.2$\cdot10^7$ & & & \citenum{gudmundsson_2007, ashida_1995} \cr
379 & $\mathrm{Ar(4p)} \to \mathrm{Ar}^m$ & $A_{ab}$ & & 3$\cdot 10^7$ & & & \citenum{gudmundsson_2007, lee_2006} \cr
380 & $\mathrm{Ar(4p)} \to \mathrm{Ar}^r$ & $A_{ab}$ & & 3$\cdot 10^7$ & & & \citenum{gudmundsson_2007, lee_2006}

 	\\
		 \br
    \end{tabular}
\end{table}
The most important transitions for VUV emission are from the $\mathrm{O(^5S)}$ and $\mathrm{O(^3S)}$ states, as they emit photons at $\sim$135 and $\sim$130\,nm when decaying to ground state.
Other transitions between excited states of oxygen atoms defined in table \ref{tab:species_list}, are included for completeness of the physical model.
However, including all possible energy transitions would add significant complexity to the collisional radiative scheme, so instead, energy transitions at higher energy levels are simplified with so-called \textit{cascade processes}\cite{fiebrandt_2020}.

Cascading processes gather several energy transition steps into one single reaction without needing to know the intermediate states.
This usually includes electron impact excitation of O atoms, or dissociative excitation during electron collisions with O$_2$ molecules, that lead to the formation of high energy levels that subsequently decay to lower energy levels that are considered as species in the numerical model.
The decay of high energy levels may occur in a stepwise manner, called cascading, and modelling this using Einstein coefficients would add significant complexity to the species and chemistry schemes.
The cascading processes incorporated to the GM are taken from Ref. \citenum{fiebrandt_2020} and are listed in table \ref{tab:oxygen_radiation}.
These processes include atomic oxygen excitation to cascading levels that results in $\mathrm{O(^5S)}$ and $\sim$777\,nm radiation, and dissociative excitation processes with O$_2$ that result in different excited oxygen atoms and radiation.
Note that the dissociative excitation processes have been extended to all O$_2$ excited states included in the GM.
%\begin{landscape}
\begin{table} %[h]
  \small
  \caption{\label{tab:oxygen_radiation}Oxygen reactions with radiative cascading processes. The reaction rates are polynomials of the form $K_r(T_e) = a_0 + a_1T_e + a_2T_e^2 + a_3T_e^3 + a_4T_e^4$ where $T_e$ is in eV.}
  \centering
    \begin{tabular}{@{}llllllll}
		 \br
		 \textbf{\#} & \textbf{Process} & \multicolumn{5}{c}{\textbf{$K_r$ [$10^{-17}$m$^{3}$s$^{-1}$]}} & \textbf{Ref.} \cr
            	&              	& $a_0$ & $a_1$ & $a_2$ & $a_3$ & $a_4$                  	& \cr
		 \mr
	%  # Table 5 ## VALUES IN THIS TABLE HAVE BEEN MULTI\mathrm{P}LIED BY 1.\cdot 10^{-6
381 & $ e + \mathrm{O} \to e + \mathrm{O}(^5\mathrm{S}) + \lambda_{777.5} $ & $-13.1$ & $52.4$ & $- 65.4$ & $29.3$ & $- 3.02$ & \citenum{fiebrandt_thesis,fiebrandt_2020,itikawa_1990}$^b$\cr
382 & $ e + \mathrm{O}_2 \to e + 2\mathrm{O} + \lambda_{130.4} $ & $-3.8 $ & $8.6 $ & $- 6.5 $ & $1.9 $ & $- 1.2 $ & \citenum{fiebrandt_thesis,fiebrandt_2020,itikawa_1990} \cr
383 & $ e + \mathrm{O}_2(\mathrm{a^1}\Delta_u) \to e + 2\mathrm{O} + \lambda_{130.4} $ & $-3.8 $ & $8.6 $ & $- 6.5 $ & $1.9 $ & $- 1.2 $ & \citenum{fiebrandt_thesis,fiebrandt_2020,itikawa_1990}$^a$ \cr
384 & $ e + \mathrm{O}_2(\mathrm{b^1}\Sigma_u^+) \to e + 2\mathrm{O} + \lambda_{130.4} $ & $-3.8 $ & $8.6 $ & $- 6.5 $ & $1.9 $ & $- 1.2 $ & \citenum{fiebrandt_thesis,fiebrandt_2020,itikawa_1990}$^a$ \cr
% 370 & $ e + \mathrm{O}_2(\mathrm{a^1}\Delta_u\,nu) \to e + 2\mathrm{O} + \lambda_{130.4}            $ & $-3.8 $ & $8.6 $ & $- 6.5 $ & $1.9 $ & $- 1.2 $   & \citenum{fiebrandt_thesis,fiebrandt_2020,itikawa_1990}$^a$ \cr
% 371 & $ e + \mathrm{O}_2(\mathrm{b^1}\Sigma_u^+,\nu) \to e + 2\mathrm{O} + \lambda_{130.4}          $ & $-3.8 $ & $8.6 $ & $- 6.5 $ & $1.9 $ & $- 1.2 $   & \citenum{fiebrandt_thesis,fiebrandt_2020,itikawa_1990}$^a$ \cr
385 & $ e + \mathrm{O}_2 \to e + 2\mathrm{O} + \lambda_{135.6} $ & $-9.20$ & $21.2$ & $- 16.5$ & $4.87$ & $- 0.331$ & \citenum{fiebrandt_thesis,fiebrandt_2020,itikawa_1990}$^b$ \cr
386 & $ e + \mathrm{O}_2(\mathrm{a^1}\Delta_u) \to e + 2\mathrm{O} + \lambda_{135.6} $ & $-9.20$ & $21.2$ & $- 16.5$ & $4.87$ & $- 0.331$ & \citenum{fiebrandt_thesis,fiebrandt_2020,itikawa_1990}$^a$$^{,b}$ \cr
387 & $ e + \mathrm{O}_2(\mathrm{b^1}\Sigma_u^+) \to e + 2\mathrm{O} + \lambda_{135.6} $ & $-9.20$ & $21.2$ & $- 16.5$ & $4.87$ & $- 0.331$ & \citenum{fiebrandt_thesis,fiebrandt_2020,itikawa_1990}$^a$$^{,b}$ \cr
% 375 & $ e + \mathrm{O}_2(\mathrm{a^1}\Delta_u\,nu) \to e + 2\mathrm{O} + \lambda_{135.6}            $ & $-9.20$ & $21.2$ & $- 16.5$ & $4.87$ & $- 0.331$  & \citenum{fiebrandt_thesis,fiebrandt_2020,itikawa_1990}$^a$$^{,b}$ \cr
% 376 & $ e + \mathrm{O}_2(\mathrm{b^1}\Sigma_u^+,\nu) \to e + 2\mathrm{O} + \lambda_{135.6}          $ & $-9.20$ & $21.2$ & $- 16.5$ & $4.87$ & $- 0.331$  & \citenum{fiebrandt_thesis,fiebrandt_2020,itikawa_1990}$^a$$^{,b}$ \cr
388 & $ e + \mathrm{O}_2 \to e + \mathrm{O} + \mathrm{O}(^5\mathrm{S}) + \lambda_{777.5} $ & $-2.91$ & $6.40$ & $- 4.67$ & $1.22$ & $- 0.0524$ & \citenum{fiebrandt_thesis,fiebrandt_2020,itikawa_1990}$^{,b}$ \cr
389 & $ e + \mathrm{O}_2(\mathrm{a^1}\Delta_u) \to e + \mathrm{O} + \mathrm{O}(^5\mathrm{S}) + \lambda_{777.5} $ & $-2.91$ & $6.40$ & $- 4.67$ & $1.22$ & $- 0.0524$ & \citenum{fiebrandt_thesis,fiebrandt_2020,mcconkey_2008}$^a$$^{,b}$ \cr
390 & $ e + \mathrm{O}_2(\mathrm{b^1}\Sigma_u^+) \to e + \mathrm{O} + \mathrm{O}(^5\mathrm{S}) + \lambda_{777.5} $ & $-2.91$ & $6.40$ & $- 4.67$ & $1.22$ & $- 0.0524$ & \citenum{fiebrandt_thesis,fiebrandt_2020,mcconkey_2008}$^a$$^{,b}$ \cr
% 380 & $ e + \mathrm{O}_2(\mathrm{a^1}\Delta_u\,nu) \to e + \mathrm{O} + \mathrm{O}(^5\mathrm{S}) + \lambda_{777.5}    $ & $-2.91$ & $6.40$ & $- 4.67$ & $1.22$ & $- 0.0524$ & \citenum{fiebrandt_thesis,fiebrandt_2020,mcconkey_2008}$^a$$^{,b}$ \cr
% 381 & $ e + \mathrm{O}_2(\mathrm{b^1}\Sigma_u^+,\nu) \to e + \mathrm{O} + \mathrm{O}(^5\mathrm{S}) + \lambda_{777.5}  $ & $-2.91$ & $6.40$ & $- 4.67$ & $1.22$ & $- 0.0524$ & \citenum{fiebrandt_thesis,fiebrandt_2020,mcconkey_2008}$^a$$^{,b}$ \cr
391 & $ e + \mathrm{O}_2 \to e + \mathrm{O} + \mathrm{O}(^3\mathrm{S}) + \lambda_{844.6} $ & $-1.55$ & $3.42$ & $- 2.51$ &$0.658$ & $- 0.0284$ & \citenum{fiebrandt_thesis,fiebrandt_2020,mcconkey_2008}$^b$ \cr
392 & $ e + \mathrm{O}_2(\mathrm{a^1}\Delta_u) \to e + \mathrm{O} + \mathrm{O}(^3\mathrm{S}) + \lambda_{844.6} $ & $-1.55$ & $3.42$ & $- 2.51$ &$0.658$ & $- 0.0284$ & \citenum{fiebrandt_thesis,fiebrandt_2020,mcconkey_2008}$^a$$^{,b}$ \cr
393 & $ e + \mathrm{O}_2(\mathrm{b^1}\Sigma_u^+) \to e + \mathrm{O} + \mathrm{O}(^3\mathrm{S}) + \lambda_{844.6} $ & $-1.55$ & $3.42$ & $- 2.51$ &$0.658$ & $- 0.0284$ & \citenum{fiebrandt_thesis,fiebrandt_2020,mcconkey_2008}$^a$$^{,b}$
% 385 & $ e + \mathrm{O}_2(\mathrm{a^1}\Delta_u\,nu) \to e + \mathrm{O} + \mathrm{O}(^3\mathrm{S}) + \lambda_{844.6}    $ & $-1.55$ & $3.42$ & $- 2.51$ &$0.658$ & $- 0.0284$ & \citenum{fiebrandt_thesis,fiebrandt_2020,mcconkey_2008}$^a$$^{,b}$ \cr
% 386 & $ e + \mathrm{O}_2(\mathrm{b^1}\Sigma_u^+,\nu) \to e + \mathrm{O} + \mathrm{O}(^3\mathrm{S}) + \lambda_{844.6}  $ & $-1.55$ & $3.42$ & $- 2.51$ &$0.658$ & $- 0.0284$ & \citenum{fiebrandt_thesis,fiebrandt_2020,mcconkey_2008}$^a$$^{,b}$

 	\\
		 \br
   \multicolumn{8}{l}{\footnotesize$^a$ This reaction is given in Ref. \citenum{fiebrandt_2020} for  electron collisions with ground state O$_2$ and is used here for electron} \\
   \multicolumn{8}{l}{\footnotesize collisions with excited states of O$_2$ with the same rate constant.} \\
   \multicolumn{8}{l}{\footnotesize$^b$ The constants were obtained from the polynomial fit to the data in the supplementary information in Ref. \citenum{fiebrandt_thesis}.} \\
    \end{tabular}
\end{table}
%\end{landscape}

Self absorption of the emission line by the lower state of the given transition can be an important effect that has an impact on the population of the emitting species and the intensity of radiation leaving the plasma.
Therefore it is important to account for this phenomena in the model.
This is modelled by adding a so called escape factor $\gamma_{ab}$, as a correction to the Einstein coefficient for spontaneous emission. To do this, we follow the approach described in \cite{fiebrandt_2020}. In general, the emission rate, $K_{ab}$, and intensity per unit volume, $I_{ab}$ for atomic transitions affected by self absorption are given by

\begin{equation}
  K_{ab} = \gamma_{ab}A_{ab}
\end{equation}

\begin{equation}
    I_{ab} = K_{ab} n_{a} \label{eqn:intensity}.
\end{equation}

%where $a$ and $b$ denote the upper and lower levels of the transition, respectively.
The definition of the escape factor used is the empirical formula given in \citenum{mewe_1967}
\begin{equation}
  \gamma_{ab} = \frac{2 - \exp{( -10^{-3} \kappa_{ab, 0} R )}}{1 + \kappa_{ab, 0} R}
\end{equation}
Under conditions where Doppler broadening is the dominant line broadening mechanism, as is the case for the low pressure conditions of interest in this work, the absorption coefficient at the centre of the emission line is given by \cite{holstein_1947}
\begin{equation}
%  \kappa_{ab} = \frac{\lambda_{ab}^2}{8\pi} P_{ab} \frac{g_p}{g_k} n_k A_{ab}
    \kappa_{ab, 0} = n_b A_{ab} \frac{g_p}{g_b} \frac{\lambda_{ab, 0}^3}{8\pi}  \sqrt{\frac{m_b}{2 k_B T_b\pi}} \label{eqn:abscoeff}
\end{equation}
where $\lambda_{ab, 0}$ is the central wavelength of the emission line.

%\begin{equation}
%  K_r = \frac{g_p}{\sum g_{p_i}} \gamma_{ab}\bigg(\frac{g_k}{\sum g_{k_i}}n_k\bigg)A_{ab}
%\end{equation}

As described in table \ref{tab:species_list}, a number of the species considered in the model consist of grouped states. While the choice to group states whose energies are very similar is convenient for the plasma-chemical model, the fact that these states emit radiation at slightly different wavelengths needs to be accounted for to properly describe the line emission and self absorption. To do this, the density distribution of individual states within a grouped state needs to be estimated. For the wavelength ranges of interest in this work, two cases can be distinguished: (1) the upper state of the transition is represented in the model by a grouped state and the lower state is not and (2) the lower state is represented in the model by a grouped state and the upper state is not. The first case applies to emission around 777\,nm (three emission lines, individual upper states: 2s$^2$ 2p$^3$ ($^3$S$^0$) 3p $^5$P$_{1,2,3}$, grouped state: O($^5$P)) and 844\,nm (three emission lines individual upper states 2s$^2$ 2p$^3$ ($^3$S$^0$) 3p $^3$P$_{1,2,0}$, grouped state: O($^3$P)). The second case applies to emission around 130 (two emission lines individual lower states 2s$^2$ 2p$^4$ $^3$P$_{2,1}$, grouped state: O) and 135\,nm (three emission lines, individual lower states 2s$^2$ 2p$^4$ $^3$P$_{2,1,0}$, grouped state: O). We follow the approach used in \citenum{fiebrandt_2020} to estimate the densities of individual multiplet states within each grouped state. Here, the density of each multiplet level is estimated using the statistical weights of each level 

%to be considered and applied as correction factor in the calculation of $\gamma_{ab}$ or $K_r$. For the optical transitions considered in this work, two cases can be distinguished. In the first case, for O atom emission at around 130 and 135 nm, the lower level of the transitions consists of multiplet components, while the upper levels do not. In these cases, the density distribution of the multiplet components in the lower level can be estimated using the statistical weights $g_k$ of the multiplet components within the lower level 

\begin{equation}
    n_{m} = \frac{g_{m}}{\sum\limits_i g_{{m}_i}} n_g \label{eqn:multiplet}
\end{equation}

where $g_m$ are the statistical weights of each multiplet level within a grouped state with density $n_g$ and $\sum\limits_i g_{m_i}$ is the sum of the statistical weights of each multiplet level within the grouped state. 

For emission around 777 and 844\,nm, where the upper state is the grouped state, the densities of the individual upper states, $n_a$, used to calculate the emission intensity in equation \ref{eqn:intensity} are determined using equation \ref{eqn:multiplet}. On the other hand, for emission around 130 and 135\,nm, where the lower state is the grouped state, the densities of the individual lower states, $n_b$ required for the calculation of $\kappa_{ab, 0}$ in equation \ref{eqn:abscoeff} are determined by equation \ref{eqn:multiplet}.

The small differences in emission wavelength of each multiplet are not relevant for the aims of the model and therefore, when presenting results, the emission intensities of the multiplet emission lines are added together.
Specifically, the 135\,nm emission line, $I_{135}$, is the sum of reactions \#366 and 367 (in table \ref{tab:radiative_transitions}), the 130\,nm line, $I_{130}$, is the sum of reactions \#367, 368 and 369, and the 777\,nm line, $I_{777}$, is the sum of reactions \#371-373.

%%%%%%%%%%%%%%%%%%%%%%%%%%%%%%%%%%%%%%%%%%%%%%%%%%%%%%%%%%%%%%%%%%%%%%%%%%%%%%%
%%%%%%%%%%%%%%%%%%%%%%%%%%%%%%%%%%%%%%%%%%%%%%%%%%%%%%%%%%%%%%%%%%%%%%%%%%%%%%%
% Experiments
\section{Experimental setup}
\label{sec:method_exp}

\begin{figure}[h]
  \centering
  \includegraphics[width=0.8\linewidth]{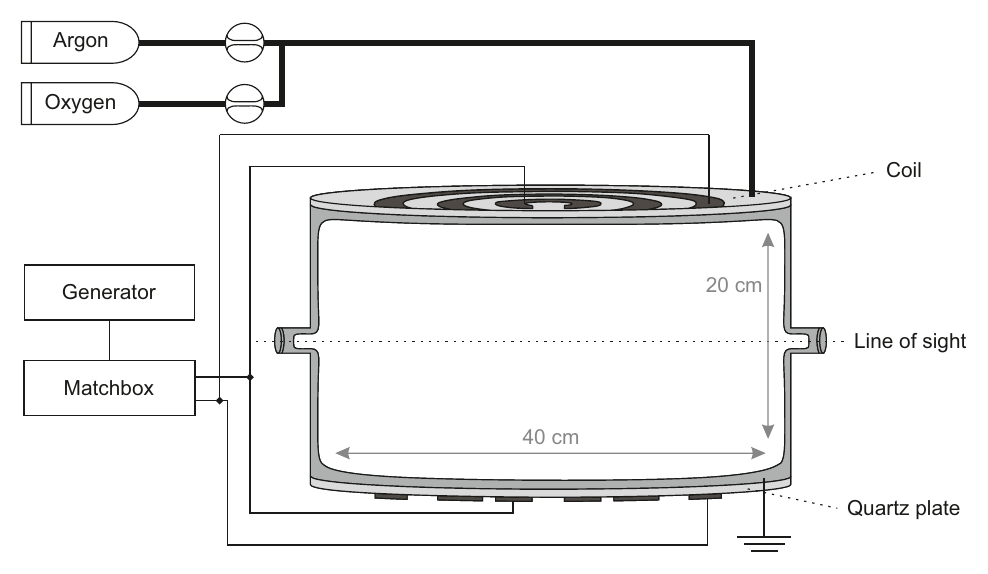}
\caption{Schematic of the DICP used for experimental validation of the simulation results.}
\label{fig:schematic_dicp}
\end{figure}

All experiments used for comparison to the simulation results were performed in a double inductively coupled plasma  (DICP) reactor as depicted in figure \ref{fig:schematic_dicp}.
The reactor comprises a cylindrical stainless steel chamber, which is $L$ = 0.2\,m in height and $R$ = 0.2\,m in radius. Several flanges are attached at half-height to allow for characterisation of plasma using optical and probe-based diagnostics. The top and bottom walls of the reactor consist of 20\,mm thick quartz plates on which the inductive coils are mounted. The generator is equipped with a matching network and operates at a driving frequency of 13.56\,MHz. Due to the reactor being powered from two sides, a relatively homogeneous plasma is obtained in the centre of the reactor. A more detailed description of the setup can be found elsewhere\cite{fiebrandt_2020, fiebrandt_2017}. For the experiments conducted in this work, a total gas flow of 100\,sccm is kept constant for all measurements. The experiments include a variation in power from 200\,W to 800\,W, a variation in pressure from 2\,Pa to 20\,Pa and a variation of the oxygen content in the Ar/O$_2$ gas mixture from 0\,\% to 20\,\%.

Measurements of electron density are conducted using a multipole resonance probe (MRP).
The MRP is based on active plasma resonance spectroscopy\cite{lapke_2013} and works by coupling an rf-signal into the plasma and measuring the response of the system. The rf-signal is varied in its bandwidth from the kHz to the GHz range, eventually inducing resonance of the electrons near the electron plasma frequency $\omega_{pe}$. Using a mathematical model, the observed resonance can be correlated to electron density $n_{e}$ and electron temperature $T_{e}$\cite{lapke_2013}. Due to the MRP relying on electron resonance, it is well suited for applications involving deposition of insulators or reactive species such as oxygen, which can affect the performance of other probe-based diagnostics. The latter is of importance here due to Ar/O$_2$ gas mixture, which would lead to deterioration of e.g. the probe tip of a Langmuir probe and consequently have the potential to lead to measurement errors. A comparison between MRP and Langmuir probe measurements is given in Fiebrandt \textit{et al} for the same setup\cite{fiebrandt_2017}.  More details on theory, operation and applications of the MRP can be found elsewhere\cite{lapke_2011,schulz_2014,pohle_2018}. Measurements require a so-called ``vacuum-trace'', which is a measurement performed without a plasma ignited for correction of conduction losses. This vacuum-trace is recorded separately for each measurement. For comparison with electron densities obtained from the GM, the probe is positioned in the centre of the setup at half-height for all measurements.

For observation of oxygen emission lines in the visible range, an echelle spectrometer ESA 4000 (LLA Instruments, Berlin) is used.
The spectrometer records spectra in the range from $\lambda =  200$\,nm to $800$\,nm and offers a resolution of between $\Delta \lambda = 0.015$\,nm and $0.06$\,nm. For calculation of absolute emission intensities, the spectrometer is absolutely calibrated as described by Bibinov \textit{et al}\cite{bibinov_2007}. Measurements are performed line-of-sight integrated at half-height in front of a quartz window. The observed plasma volume is defined by an aperture mounted on the optical fibre (acceptance angle $\theta = 1.58$\,°). Of particular interest with regards to comparing with the simulation results is the $\mathrm{O(^5P_{1,2,3})}\to \mathrm{O(^5S)}$ transition, measured at $777$\,nm.
By integrating the absolutely calibrated spectra over the emission lines from 777.07\,nm to 777.65\,nm, absolute intensities are obtained.

Tunable diode laser absorption spectroscopy (TDLAS) is performed to measure gas temperature and argon metastable densities. Specifically, the Ar(1s$_5$) metastable state is measured using the Ar(1s$_5$ $\rightarrow$ 2p$_6$) transition at 772.376\,nm.
The system employed for the measurements consists of a laser head (DFB pro 100\,mW, 772\,nm + Fiberdock) and a laser controller (DLC pro). The laser beam traverses the plasma chamber in full diameter and is detected by a photodiode (Thorlabs DET10N2). In addition to a photodiode, a fraction of the laser power is coupled to a Fabry-Perot interferometer (Toptica FPI 100-750-3V0, 1\,GHz free spectral range), allowing for monitoring of the change of the scanning laser wavelength. For each measurement point, four individual measurements are performed: (i) plasma on and laser on, (ii), plasma on and laser off, (iii), plasma off and laser on, (iv) plasma off and laser off. These four measurements are required for processing of the data. Gas temperatures and argon metastable densities are obtained by applying a Gaussian fit to the absorption profile. The gas temperature is calculated assuming that the line width, for the pressure range in this work, arises mainly from Doppler broadening, which can be directly related to the gas temperature. The calculation is performed by a semi-automatic LabVIEW software. The full setup of the TDLAS system and evaluation of the acquired data is described by Schulenberg \textit{et al}\cite{Schulenberg_2021}.

%%%%%%%%%%%%%%%%%%%%%%%%%%%%%%%%%%%%%%%%%%%%%%%%%%%%%%%%%%%%%%%%%%%%%%%%%%%%%%%
%%%%%%%%%%%%%%%%%%%%%%%%%%%%%%%%%%%%%%%%%%%%%%%%%%%%%%%%%%%%%%%%%%%%%%%%%%%%%%%
%%%%%%%%%%%%%%%%%%%%%%%%%%%%%%%%%%%%%%%%%%%%%%%%%%%%%%%%%%%%%%%%%%%%%%%%%%%%%%%
% Results
\section{Results}
\label{sec:results}

\subsection{Characterization of Ar/O$_2$ DICP with numerical and experimental data}
\label{sec:results_part1}
The influence of variations of total pressure $p_T$, power $P_{in}$ and oxygen gas fraction $\chi_{\mathrm{O}_2}$ on the plasma properties are presented.
The total pressure is varied between $p_T=$ 2 - 20\,Pa the input power $P_{in}=$ 200 - 800\,W and the oxygen fraction $\chi_{\mathrm{O}_2}$ = 0 - 0.20.
However, since the temperature of ions and neutrals, $T_N$, changes significantly under variations of $p_T$, $P_{in}$, and $\chi_{\mathrm{O}_2}$\cite{fiebrandt_2017}, and this is a fixed parameter in the GM, simulations are run with various values of $T_N$ to ensure that variations of this parameter have been taken into account in the final results.
On the one hand, simulations have been performed varying $T_N$ between 400 - 2000K in order to understand the impact of $T_N$ on the plasma parameters.
\begin{table}[h]
  \caption{\label{tab:T_N}Neutral gas temperature experimental measurements, in K. The error shows the standard deviation obtained from three measurements for each operating condition.}
  \centering
    \begin{tabular}{@{}l | lll | lll}
		 \br
		 \multirow{2}{2.5em}{\textbf{$\chi_{\mathrm{O}_2}$}} & \multicolumn{3}{c|}{\textbf{5 Pa}}& \multicolumn{3}{c}{\textbf{500 W}} \\
		      	& 200 W & 500 W & 800 W & 2 Pa & 10 Pa & 20 Pa \\
		 \mr
 	        0.0 &  425$\pm$11  & 513$\pm$ 8 & 569$\pm$ 3 & 413$\pm$55 & 632$\pm$ 5 & 787$\pm$12 \\
 	        0.04 & 567$\pm$11  & 657$\pm$ 6 & 722$\pm$16 & 459$\pm$ 5 & 680$\pm$10 & 780$\pm$ 9 \\
 	        0.08 & 615$\pm$16  & 743$\pm$12 & 843$\pm$49 & 501$\pm$ 8 & 675$\pm$31 & 446$^a$ \\
 	        0.12 & 626$\pm$36  & 801$\pm$10 & 862$\pm$10 & 538$\pm$ 7 & 654$\pm$22 & 587$\pm$68$^b$ \\
 	        0.16 & 617$\pm$35  & 784$\pm$12 & 931$\pm$25 & 534$\pm$ 7 & 661$\pm$12 & \\
 	        0.2  & 596$\pm$ 7  & 793$\pm$ 9 & 930$\pm$ 1 & 526$\pm$22 & 726$\pm$22$^b$ &
 	\\
		 \br
   \multicolumn{7}{l}{\footnotesize$^a$ One valid measurement was taken.} \cr
   \multicolumn{7}{l}{\footnotesize$^b$ Two valid measurements were taken.} \cr
    \end{tabular}
\end{table}
On the other hand, a second set of simulations has also been run using values of $T_N$ measured experimentally using TDLAS, listed in table \ref{tab:T_N}, in order to better compare experiment and simulation.
The results are compared with the experimental work described in section \ref{sec:method_exp} and with results from Fiebrandt \textit{et al} in \citenum{fiebrandt_2017}, \citenum{fiebrandt_2017_1} and \citenum{fiebrandt_2020}.
The experimental work from Fiebrandt \textit{et al} is conducted on the same plasma reactor and in similar operating conditions and thus its results are a useful reference. However, in the time since the earlier works of Fiebrandt \textit{et al}, the reactor has undergone several changes including the replacement of the quartz plates separating the coils from the plasma. While these changes would not necessarily be expected to significantly affect the plasma properties, since the design of the reactor has not changed, the more recent measurements are generally not in exact agreement with the earlier data for otherwise identical operating conditions. This should also be kept in mind when interpreting the level of agreement between experiment and simulation.
Therefore, the results  presented in this section are not only used to provide a general characterization of Ar/O$_2$ plasmas and its radiation behaviour of oxygen species, but also to validate the numerical results.

\subsubsection{Electron density and temperature}
\begin{figure}[h]
  \centering
  \includegraphics[width=\linewidth]{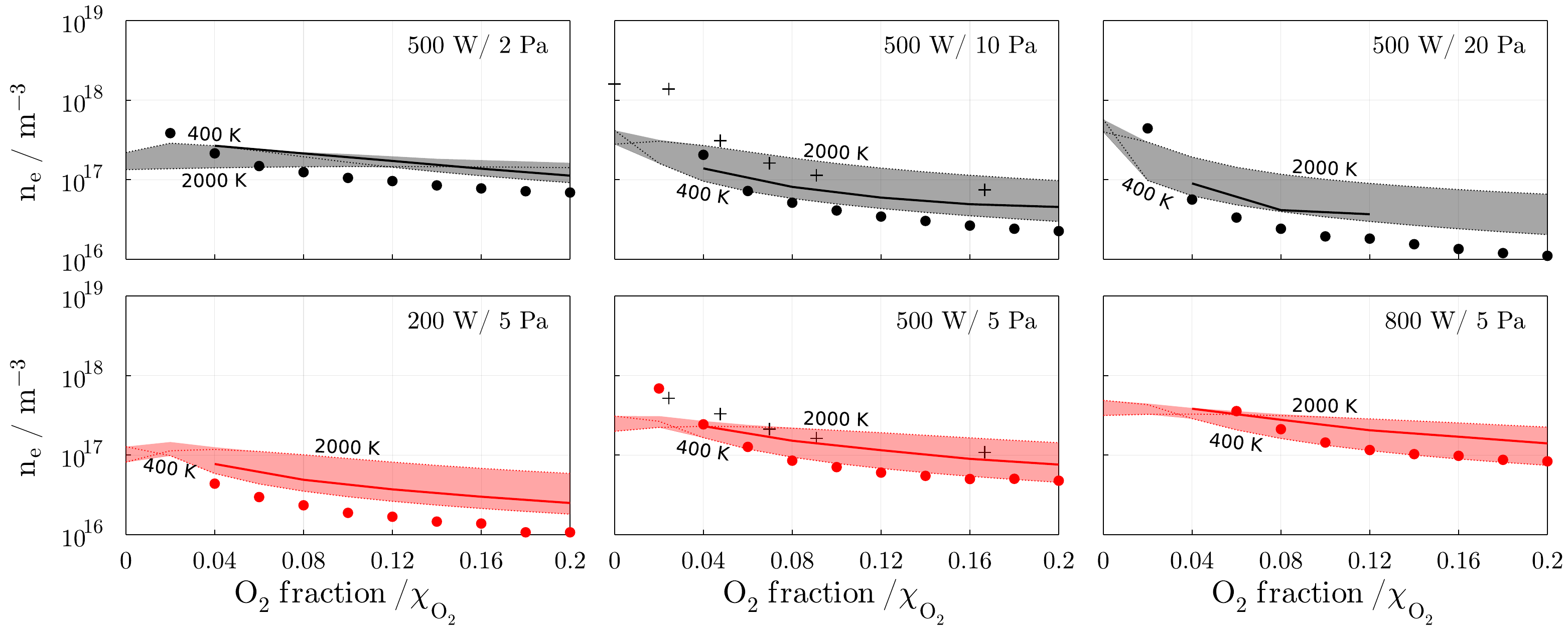}
\caption{\label{fig:ne}Electron density, $n_e$, for variations of $p_T$ (top row), $P_{in}$ (bottom row) and $\chi_{\mathrm{O}_2}$. Circle ($\bullet$) markers are experimental data described in section \ref{sec:method_exp}, and $+$ markers are experimental MRP results from \citenum{fiebrandt_2017_1}. The shaded areas cover the model results when the neutral gas temperature, $T_n$, is varied between 400 and 2000\,K (dotted lines). The solid lines are numerical results using the $T_N$ experimental data listed in table \ref{tab:T_N}.}
\end{figure}

\begin{figure}[h]
  \centering
  \includegraphics[width=0.5\linewidth]{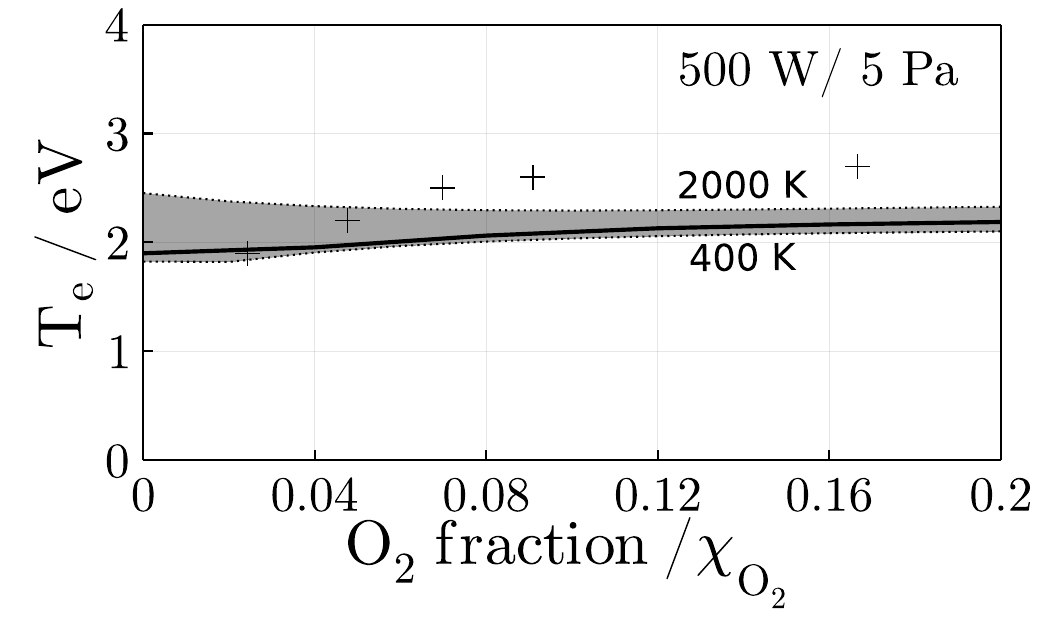}
\caption{\label{fig:Te}Electron temperature, $T_e$, for variations of O$_2$ fraction.
The $+$ markers are LP experimental results in \citenum{fiebrandt_2017_1}. The shaded areas cover the model results when the neutral gas temperature, $T_n$, is varied between 400 and 2000\,K (dotted lines). The solid lines are numerical results using the $T_N$ experimental data listed in table \ref{tab:T_N}.}
\end{figure}

%ELECTRON DENSITY
The electron density $n_e$, in figure \ref{fig:ne}, and temperature $T_e$, in figure \ref{fig:Te}, are the first parameters to evaluate the plasma results.
The numerical results for $n_e$ are compared with MRP measurements described in section \ref{sec:method_exp} and also conducted in \citenum{fiebrandt_2017_1}.
Numerical results for $T_e$ are compared with Langmuir probe data from \citenum{fiebrandt_2017_1}.

First, $n_e$ values are found between $10^{16}$ and $10^{18}$\,m$^{-3}$ and present decreasing trends with $\chi_{\mathrm{O}_2}$, as observed in \citenum{gudmundsson_2007}, as well as with $p_T$.
These trends are caused by a constant growth of dissociative attachment (reactions \#25, 42 and 60 in table \ref{tab:electron_oxygen}) acting as the main electron loss mechanism, while the main production mechanism transits from argon ionization (reaction \#108 in table \ref{tab:electron_argon}), dominant at low  $\chi_{\mathrm{O}_2}$ and $p_T$, to the recombination of O$_2((\mathrm{a}^1\Delta_u)$ and O$_2(\mathrm{b}^1\Sigma_u^+)$ with O$_2^-$ (reactions \#227 and 231 in table \ref{tab:oxygen_oxygen}), and O with O$^-$ (reaction \#177 in table \ref{tab:oxygen_oxygen}), at low  $\chi_{\mathrm{O}_2}$ and higher $p_T$,

Besides, a positive trend in $n_e$ with $P_{in}$ is observed that is in line with the results in \citenum{gudmundsson_1999}. This is caused by a significant increase of argon ionization with increasing $P_{in}$.

The simulation results and experimental data are in good agreement with both showing similar trends for variations of $p_T$, $P_{in}$ and $\chi_{\mathrm{O}_2}$.
There is however a consistent difference between numerical and experimental results (circle markers), with the latter generally being slightly lower.
A potential explanation for this may lie in the fact that the power defined for the simulation is that coupled into the plasma, that defined for the experiment is measured at the RF generator.
It is generally well known that there can be significant differences between the power provided at the RF generator and the power coupled into the plasma in ICP systems\cite{godyak_2011,zielke_2021,rauner_2019,elsaissi_2022, rauner_2017,schwabedissen_1997}.
Since the electron density is strongly power dependent, any deviation between generator power and that coupled into the plasma would tend to decrease the experimentally measured electron density in comparison to the simulated electron density.
However, since we are currently unable to characterise the power coupling efficiency in detail, the extent to which this effect can explain the differences between experiment and simulation is currently not known.

% ELECTRON TEMPERATURE
The electron temperature data $T_e$, in figure \ref{fig:Te}, show values between approximately 2 and 3\,eV for variations in $\chi_{\mathrm{O}_2}$.
Both numerical and experimental values, show a slight positive trend that plateaus with increasing values of $\chi_{\mathrm{O}_2}$.
The absolute $T_e$ values between experimental and numerical data differs less than 1\,eV and therefore results are in reasonably good agreement.
The assumption of a Maxwellian electron EDF, which does not hold for increasing $\chi_{\mathrm{O}_2}$\cite{gudmundsson_1999, lee_2013,fiebrandt_2017_1}, is likely to be an important reason for the differences that do exist between experiment and simulation.
While this is a weakness in the model formulation, the effect on the comparison between experimentally measured and simulated electron densities and temperatures is not severe for the cases compared here. A detailed study on the effects of the EDF shape on the properties of oxygen discharges for similar conditions has previously been carried out in \citenum{toneli_2015}. In general, EDFs of different shapes were found to change the absolute values of species densities and electron temperatures predicted by the global model used in that work, without strongly affecting the observed trends. 
Given this context and the $n_e$ and $T_e$ comparisons obtained here it can be concluded that the physics and chemistry modelled by the GM is as expected and is in good agreement with experimental work and previous literature.

\subsubsection{Role of neutral gas temperature}
% NEUTRAL TEMPERATURE VARIATIONS
The variations of $n_e$ and $T_e$ caused by variations of $T_N$, shown in the figures by the shaded areas, are considerable but do not have a determining effect on the trends observed.
The resulting plasma parameters remain within an order of magnitude for variations between 400 and 2000\,K.
Similar variations are observed for the other parameters described in this section, so it can be concluded that $T_N$ has an important influence on the plasma properties, but does not have a strong influence on the qualitative trends presented in this work.

\subsubsection{Neutral species densities}
With respect to neutral species densities, measured and simulated densities of Ar$^m$, $n_{\mathrm{Ar}^m}$, and the O$_2$ dissociation fraction are compared.

In model, the species Ar$^m$ represents an effective metastable state that includes the states Ar($1s_3$) and Ar($1s_5$)\cite{gudmundsson_2007}.
However, the experimental measurements performed with TDLAS, described in section \ref{sec:method_exp}, measure only the Ar($1s_5$) state. 
Still, the comparison between the experimental and GM results is considered reasonable since the work performed in \citenum{fiebrandt_2017} with optical emission spectroscopy (OES), under similar operating conditions, infers the densities of both Ar($1s_3$) and Ar($1s_5$) states and shows that the former is typically an order of magnitude lower in density.

%These parameters condense the main characteristics of the gas and are therefore considered sufficient for its description and numerical validation.

\begin{figure}[h]
  \centering
  \includegraphics[width=\linewidth]{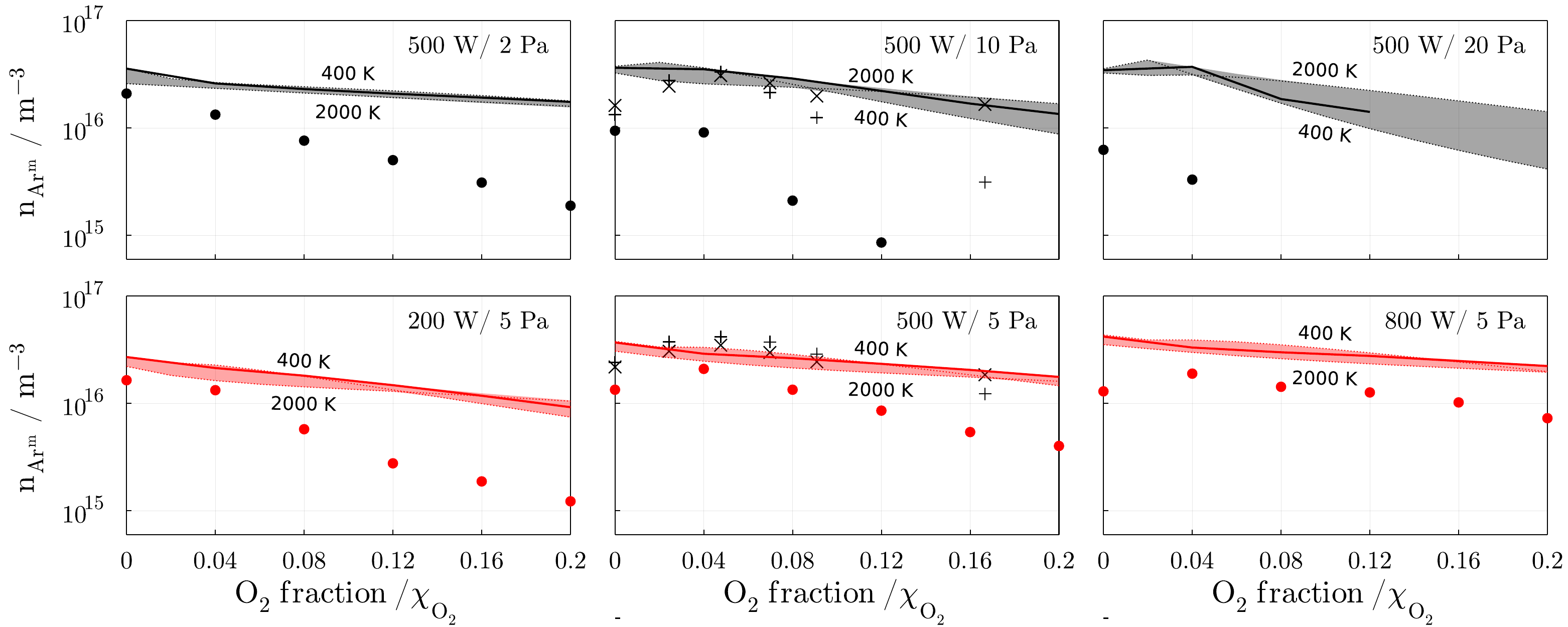}
\caption{\label{fig:Arm_power}A$r^m$ density, $n_{\mathrm{Ar}^m}$, for variations of $p_T$ (top row), $P_{in}$ (bottom row) and $\chi_{\mathrm{O}_2}$.
Circle ($\bullet$) markers are experimental TDLAS data described in section \ref{sec:method_exp}, $+$ and $\times$ markers are TDLAS and OES results in \citenum{fiebrandt_2017}, respectively. The shaded areas cover the model results when the neutral gas temperature, $T_n$, is varied between 400 and 2000\,K (dotted lines). The solid lines are numerical results using the $T_N$ experimental data listed in table \ref{tab:T_N}. It should be noted that TDLAS measurements refer to the density of the Ar(1s$_5$) state, while the simulated densities and OES measurements represent an effective metastable state comprising the densities of both Ar(1s$_3$) and Ar(1s$_5$).}
\end{figure}

% ARGON-M
The results for $n_{\mathrm{Ar}^m}$, in figure \ref{fig:Arm_power}, show values between 10$^{15}$ and 10$^{17}$\,m$^{-3}$.
The production of Ar$^m$ is sustained by electron impact excitation from ground and radiative Ar$^r$ states, reactions \#109-110 and 123 in table \ref{tab:electron_argon} respectively, and the decay Ar(4p)$\to$Ar$^m$, reaction \#379 in table \ref{tab:radiative_transitions}. These three reactions are of similar importance in the range of parameters studied.
The loss mechanisms of Ar$^m$ are dominated by electron impact collisions forming Ar$^r$ and Ar(4p) (\#116-117 in table \ref{tab:electron_argon}), and the dissociation of O$_2$ by Ar$^m$ impact (\#276 and 282 in table \ref{tab:argon_oxygen}), which is expected to be important when $\chi_{\mathrm{O}_2}\to 1$\cite{agarwal_2003, takechi_2001, kitajima_2006, gudmundsson_2007}, is only relevant for $P_{in}$ = 200\,W and $\chi_{\mathrm{O}_2}\simeq 0.2$.

The GM results and the experimental measurements carried out in this work (circle markers) show reasonable agreement as they share similar trends and results are, mostly, within an order of magnitude in terms of absolute values.
The differences between GM and experimental work become more pronounced for increasing $p_T$ and $\chi_{\mathrm{O}_2}$.
The reason for these divergences are not fully clear as there are many factors that could be involved, both from the experimental and the computational perspectives.
On the experimental side, note that measurements carried out in \citenum{fiebrandt_2017} using TDLAS and OES, $+$ and $\times$ markers respectively in figure \ref{fig:Arm_power}, show better agreement with the GM results than the measurements done in this investigation. This may reflect changes in the experimental system between now and when the work of Fiebrandt was carried out, as discussed earlier.
On the simulation side, the GM results are consistently above the experimental data, as also observed for $n_e$ in figure \ref{fig:ne}, and therefore a discrepancy with the experimental data due to a non-unity inductive power coupling efficiency cannot be discarded.

%Further Ar$^m$ results found in the literature\cite{gudmundsson_2007,sato_2008,hayashi_1999,boffard_2014_2} are consistent with the results shown in figure \ref{fig:Arm_power}, i,e. results are within an order of magnitude and showing similar trends.
Aside from comparing with experimental data, a series of simulations has also been carried out to compare with previous simulations of Ar excited state densities in Ar/O$_2$ plasmas with varying O$_2$ content\cite{gudmundsson_2007,sato_2008}. In general, very good agreement (not shown) is found in the excited state densities of Ar simulated in those previous works and using the GM developed here.
%The results of these simulations, shown and described in \ref{app:sato}, \ref{app:gudmundsson} and \ref{app:hayashi}, are in good agreement with their respective references and we therefore consider the numerical results of $n_{Ar^m}$ to be acceptable.

% OXYGEN dissociation percentages
\begin{figure}[h]
  \centering
  \includegraphics[width=0.5\linewidth]{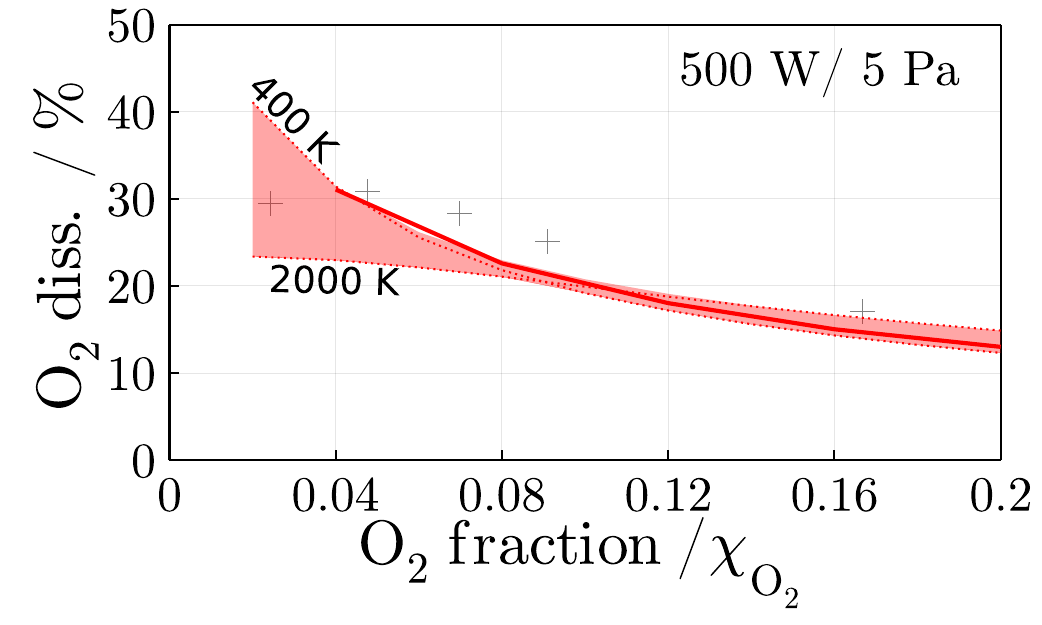}
\caption{\label{fig:Odiss}Oxygen dissociation percentage for variations of $\chi_{\mathrm{O}_2}$.
The $+$ markers are the collisional-radiative model results in \citenum{fiebrandt_2020}. The shaded areas cover the model results when the neutral gas temperature, $T_n$, is varied between 400 and 2000\,K (dotted lines). The solid lines are numerical results using the $T_N$ experimental data listed in table \ref{tab:T_N}.}
\end{figure}
The oxygen dissociation percentage
\begin{equation}
  \mathrm{O_2\,diss.[\%]} = 100 \frac{\frac{1}{2}n^*_\mathrm{O}}{\frac{1}{2}n^*_\mathrm{O} + n^*_{\mathrm{O_2}}},
\end{equation}
where $n^*_\mathrm{O}$, and $n^*_{\mathrm{O_2}}$, are the sum of all atomic, and molecular, oxygen species in table \ref{tab:species_list}, reflecting the ratio between atomic and molecular oxygen present in the system.
The dissociation percentage are shown in figure \ref{fig:Odiss}, where GM results are compared to the collisional-radiative model (CRM) results in \citenum{fiebrandt_2020}.
The CRM estimates volume averaged atomic oxygen ground and excited state densities from experimental data.
Both CRM and GM results are in good agreement, showing a decreasing trend for growing $\chi_{\mathrm{O}_2}$.
This shows that GM results for the main oxygen species, i.e. the molecular and atomic species in the ground state, are computed as expected.

\subsubsection{Oxygen radiation}
\label{sec:results_oxygen_radiation}
The simulation of radiation from oxygen species is tested with the 777\,nm emission line, $I_{777}$, from the $\mathrm{O(^5P)} \to \mathrm{O(^5S)}$ transition, and the VUV emission lines, $I_{VUV} = I_{130} + I_{135}$.
The two most important VUV emission lines investigated are the 130\,nm line, $I_{130}$, from the $\mathrm{O(^3S)}\to \mathrm{O}$ transition, and the 135\,nm line, $I_{135}$, from the $\mathrm{O(^5S)} \to \mathrm{O}$ transition.
These parameters are not only used to study the radiation of oxygen but also to verify the composition of excited states present in the gas.

\begin{figure}[h]
  \centering
  \includegraphics[width=\linewidth]{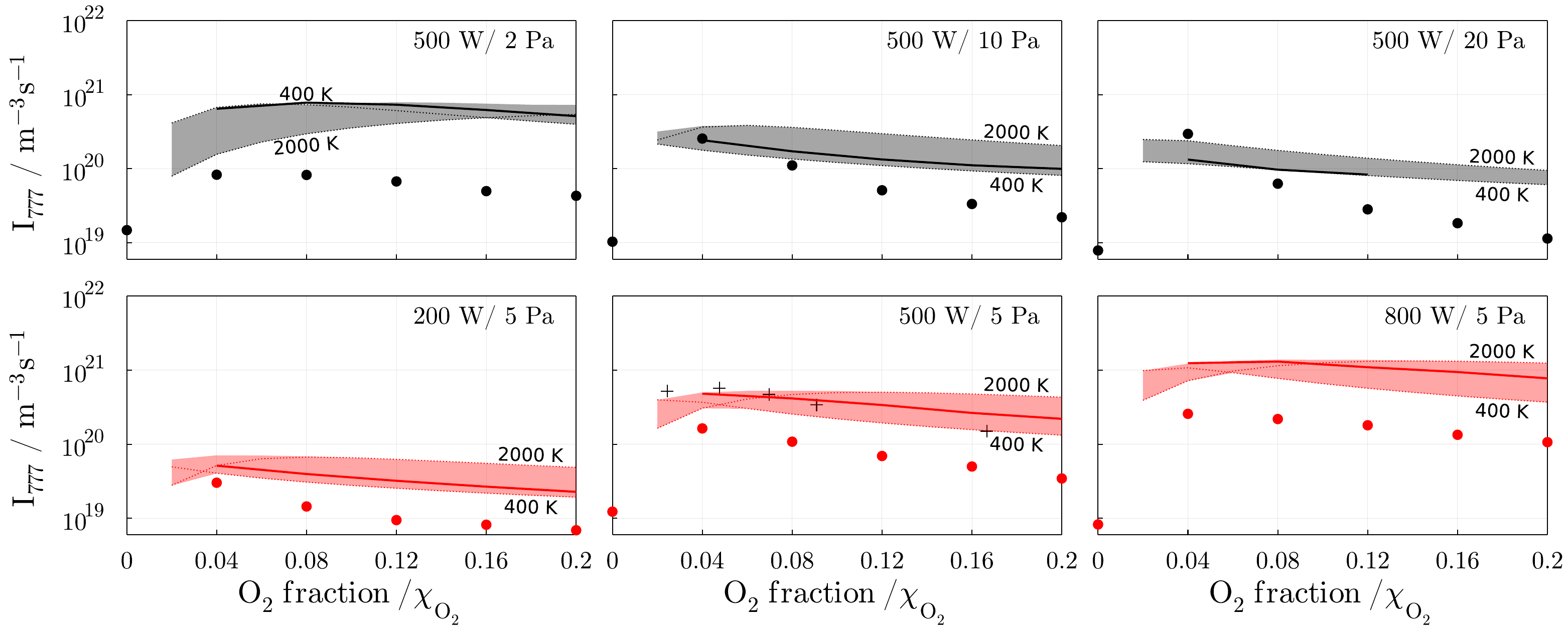}
\caption{\label{fig:I777nm}Emission intensity of the 777\,nm line, from transition $\mathrm{O(^5P)}\to \mathrm{O(^5S)}$, for variations of $p_T$ (top row), $P_{in}$ (bottom row) and $\chi_{\mathrm{O}_2}$. Circle ($\bullet$) markers are experimental spectrometer data described in section \ref{sec:method_exp}, and $+$ spectromenter results in Refs. \citenum{fiebrandt_2018, fiebrandt_2020}. The shaded areas cover the model results when the neutral gas temperature, $T_n$, is varied between 400 and 2000\,K (dotted lines). The solid lines are numerical results using the $T_N$ experimental data listed in table \ref{tab:T_N}.
}
\end{figure}

% I 777nm EMISSION LINE
The results for $I_{777}$, in figure \ref{fig:I777nm}, show emission intensities between 10$^{19}$ and 10$^{21}$\,m$^{-3}$s$^{-1}$.
The production of O($^5P$) is mostly sustained by electron impact excitation from ground state (reaction \#88 in table \ref{tab:electron_oxygen}) and from O($^5S$) (\#99  in table \ref{tab:electron_oxygen}), where the latter is more important when $P_{in}$ is larger, $p_T$ is lower, and/or $\chi_{\mathrm{O}_2}\to 0$. 
The main loss mechanism of O($^5P$) is the decay $\mathrm{O(^5P)}\to \mathrm{O(^5S)}$ (\#371-373 in table \ref{tab:radiative_transitions}) that emits at 777\,nm.
Although O($^5P$) is directly responsible for the 777\,nm line, the concentration of O($^5S$) is also important as it is closely related to the creation and destruction of O($^5P$).
As expected, O($^5S$) is mainly created by electron impact excitation (reaction \#86 in table \ref{tab:electron_oxygen}) and the transition $\mathrm{O(^5P)}\to \mathrm{O(^5S)}$. However, the destruction of  O($^5S$) is not only determined by electron impact excitation to O($^3P$),  O($^3S$) and  O($^5P$) but also by quenching with Ar, O and O$_2$. Quenching reactions become more important at increasing $p_T$ and $\chi_{\mathrm{O}_2}$ and are thus responsible for the decreasing trends with respect to these parameters.

The results obtained with the GM are in reasonably good agreement with experimental measurements carried out in this work, as trends are similar and values differ less than an order of magnitude.
%The experimental results in \citenum{fiebrandt_2018}, at 5\,Pa and 500\,W, show very good agreement.
The experimental data conducted in this investigation is systematically below the numerical data, and that of the previous work of Fiebrandt\cite{fiebrandt_2018, fiebrandt_2020}, as observed above for $n_e$ and $n_{Ar^m}$.
Although it is not yet clear what the cause of this difference is, the low power coupling efficiency could be an important factor to take into account, as the coupling efficiency decreases with low pressure and high power\cite{godyak_2011}, and this is in consistent with the observed differences between the experimental and numerical results.
However, other factors must also be taken into account for the deviation between numerical and experimental data.
%For example, $I_{777}$ depends on the density of $\mathrm{O(^5P)}$, which is about 7-8 orders of magnitude lower than that of atomic and molecular oxygen, so the accuracy and completeness of the reaction scheme plays an important role, as well as the assumptions made in the model.
Therefore, bearing in mind the simplifications made, the results of the GM are taken as acceptable.

\begin{figure}
  \centering
  \includegraphics[width=0.5\linewidth]{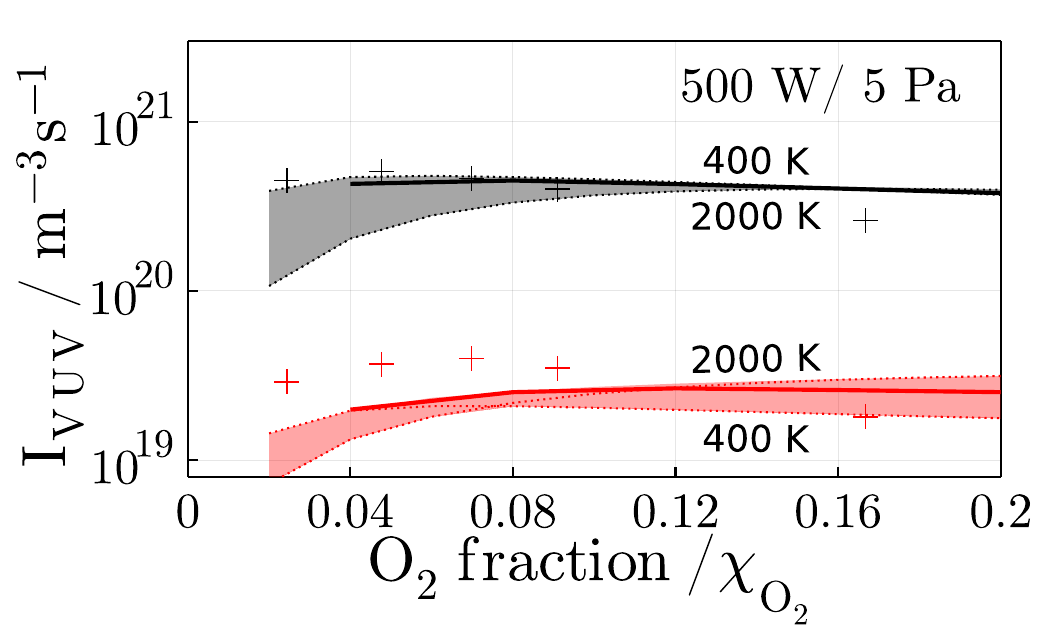}
\caption{\label{fig:VUV_Tneutral}VUV emission intensities for variations of $\chi_{\mathrm{O}_2}$.
In black, the 130\,nm line transition $\mathrm{O(^3S)}\to \mathrm{O}$ and, in red, the 135 nm line transition $\mathrm{O(^5S)}\to \mathrm{O}$.
The $+$ markers are the results in \citenum{fiebrandt_2020}. The shaded areas cover the model results when the neutral gas temperature, $T_n$, is varied between 400 and 2000\,K (dotted lines). The solid lines are numerical results using the $T_N$ experimental data listed in table \ref{tab:T_N}.}
\end{figure}

%VUV EMISSION
The VUV emission results, shown in figure \ref{fig:VUV_Tneutral}, show good agreement between the GM results and the experimental data in \citenum{fiebrandt_2020}..
The 130\,nm emission line, $I_{130}\sim$ $5\cdot10^{20}$m$^{-3}$s$^{-1}$, dominates the oxygen VUV radiation as it is an order of magnitude higher than the 135\,nm line, $I_{135}\sim$ $5\cdot10^{19}$m$^{-3}$s$^{-1}$.
For both emission lines, radiation comes from the natural decay of excited species, $\mathrm{O(^3S)}\to \mathrm{O}$ (reactions \#368-370 in table \ref{tab:radiative_transitions}) and $\mathrm{O(^5S)}\to \mathrm{O}$ (reactions \#366-367 in table \ref{tab:radiative_transitions}) respectively, and the contribution from cascading reactions (\#382-387 in table \ref{tab:oxygen_radiation}) is negligible.
This is in line with the description given in \citenum{boffard_2014_2}.
Further analysis of oxygen VUV radiation is found in the following section.

%%%%%%%%%%%%%%%%%%%%%%%%%%%%%%%%%%%%%%%%%%%%%%%%%%%%%%%%%%%%%%%%%%%%%%%%%%%%%%%%
%%%%%%%%%%%%%%%%%%%%%%%%%%%%%%%%%%%%%%%%%%%%%%%%%%%%%%%%%%%%%%%%%%%%%%%%%%%%%%%%
%%%%%%%%%%%%%%%%%%%%%%%%%%%%%%%%%%%%%%%%%%%%%%%%%%%%%%%%%%%%%%%%%%%%%%%%%%%%%%%%
\subsection{Vacuum ultraviolet emission in oxygen species}
\label{sec:results_part2}
After confirming that GM results are in good agreement with experimental reality, this second part of the results presents an extended numerical investigation of VUV radiation in Ar/O$_2$ plasmas.
The results over a wider range of operating conditions, $P_{T}=$ 0.3-100\,Pa and $p_T=$ 100-2000\,W, are presented and analysed.
The analysis of the results focuses on the VUV emission intensity of oxygen species, in absolute terms, $I_{VUV}$, but also with respect to the flux of ions, $\frac{V}{A}I_{VUV}/\Gamma_+$, and oxygen atoms, $I_{VUV}/R_{D,\mathrm{O}}$, present in the DICP system, as these are quantities that are generally known to be important for the understanding and optimisation of various surface treatments.

%%%%%%%%%%%%%%%%%%%%%%%%%%%%%%%%%%%%%%%%%%%%%%%%%%%%%%%%%%%%%%%%%%%%%%%%%%%%%%%%
%%%%%%%%%%%%%%%%%%%%%%%%%%%%%%%%%%%%%%%%%%%%%%%%%%%%%%%%%%%%%%%%%%%%%%%%%%%%%%%%
\subsubsection{Absolute VUV emission intensities}
\label{sec:VUV_absolute}
\begin{figure}[h]
  \centering
  \includegraphics[width=\linewidth]{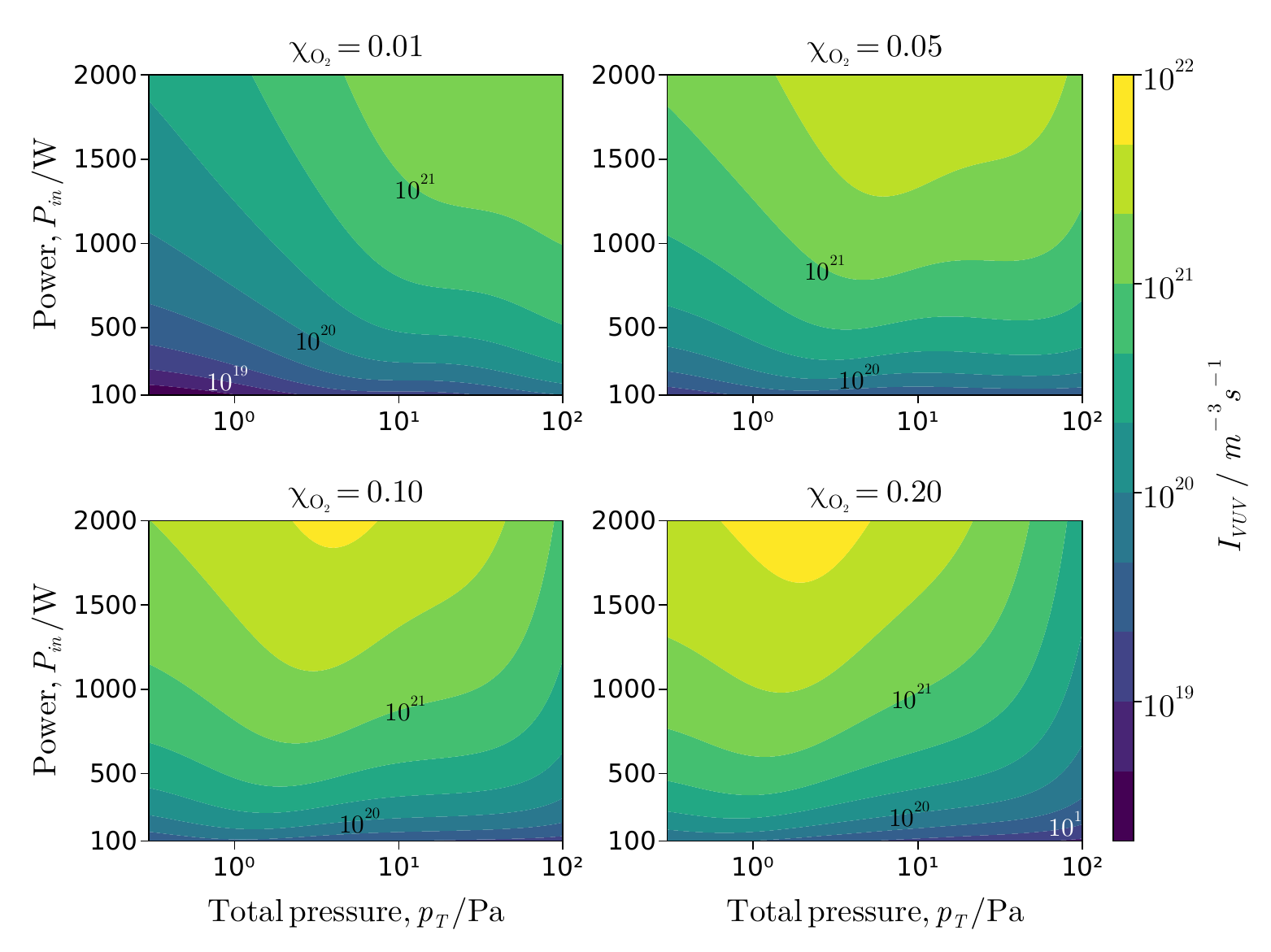}
\caption{\label{fig:VUV_power_press}Absolute vacuum-ultraviolet (VUV) emission intensity, $I_{VUV}$, from oxygen species for variations of $p_T$, $P_{in}$, and $\chi_{\mathrm{O}_2}$.}
\end{figure}

The total VUV emission intensity from oxygen species, $I_{VUV}$, is shown in figure \ref{fig:VUV_power_press}.
These results show that, in general terms, the VUV radiation is higher at higher $P_{in}$ and $\chi_{\mathrm{O}_2}$ and finds a peak at a given range of $p_T$. %as these parameters contribute to a larger presence of atomic oxygen necessary for VUV radiation.
This VUV peak with respect to $p_T$ moves towards lower pressure values as the $\chi_{\mathrm{O}_2}$ increases.
The VUV emission, as noted in section \ref{sec:results_oxygen_radiation} is dominated by the 130\,nm line, specifically by the transition $\mathrm{O(^3S)}\to \mathrm{O}$.
%Similar behaviour with pressure is observed for Ar VUV emission in \citenum{boffard_2014_1, tian_2015}.

%\begin{figure}[h]
%  \centering
%  \includegraphics[width=\linewidth]{figures/Odiss_power_press.pdf}
%\caption{\label{fig:Odiss_power_press}$\mathrm{O_2}$ dissociation \% percentage.}
%\end{figure}
%In fact, the VUV radiation from oxygen can be classified into two clearly distinct regions depending on the dissociation percentage.
%For high O$_2$ dissociation percentages, i.e. $\ge50\%$, the VUV emission is mainly dominated by the 130\,nm line, as observed in figure \ref{fig:VUV_Tneutral}, and the 135\,nm line an order of magnitude lower and thus negligible.
%At high dissociation percentages, atomic oxygen is abundant, and thus the decay of excited species dominates the VUV radiation, especially the $\mathrm{O(^3S)} \to \mathrm{O}$ transition responsible for the 130\,nm light.
%At low dissociation percentages, $\le10\%$ at low $P_{in}$ and high $p_T$, the absolute VUV emission intensities are low due to the absence of atomic oxygen and the cascading becomes the main VUV radiation process.
%In this case, the 130 and 135\,nm lines are equally important, each accounting for about 40\% of the total emission.

\begin{figure}[h]
  \centering
  \includegraphics[width=\linewidth]{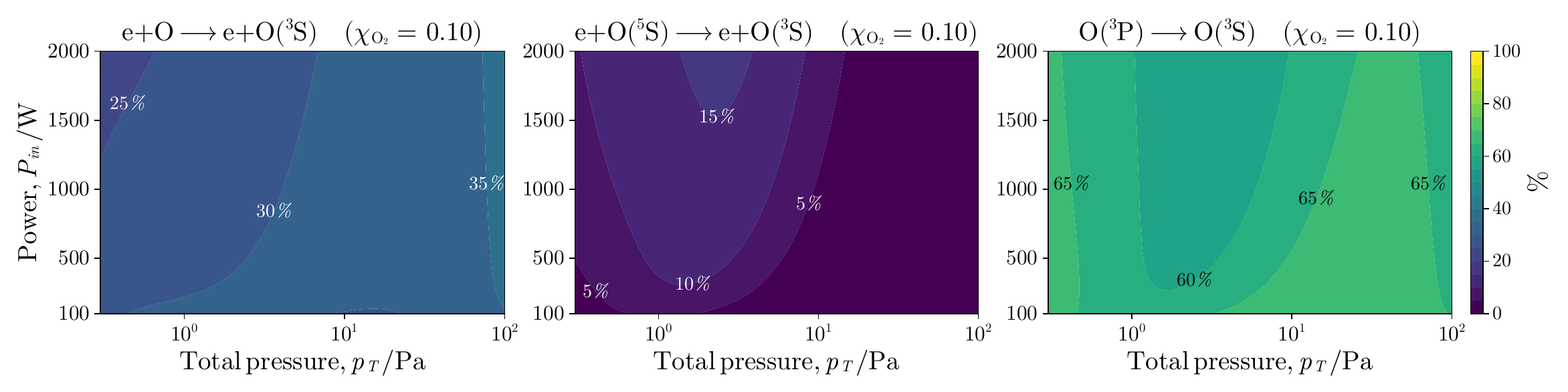}
\caption{Most important $\mathrm{O(^3S)}$ production processes as a \% of the overall $\mathrm{O(^3S)}$ production for $\chi_{\mathrm{O}_2} = 0.1$. }
\label{fig:O3s_pathways_power_press}
\end{figure}

The reaction pathways for the production of $\mathrm{O(^3S)}$ species have been tracked to understand the most important source of oxygen VUV radiation.
The main production mechanisms of $\mathrm{O(^3S)}$ are electron impact excitation of atomic oxygen $e + \mathrm{O} \to e + \mathrm{O(^3S)}$ (reaction \#87 in table \ref{tab:electron_oxygen}), electron impact cross-excitation $e + \mathrm{O(^5S)} \to e + \mathrm{O(^3S)}$ (\#98 in table \ref{tab:electron_oxygen}), and the radiative decay $\mathrm{O(^3P)} \to \mathrm{O(^3S)}$ (\#374-376 in table \ref{tab:radiative_transitions}).
The \% of $\mathrm{O(^3S)}$ produced by each of these reactions is shown in figure \ref{fig:O3s_pathways_power_press} for the case where $\chi_{\mathrm{O}_2}=0.1$.
Interestingly, the most frequent production mechanism of $\mathrm{O(^3S)}$ is via decay $\mathrm{O(^3P)}$, about 60-70\%, instead of the direct excitation through electron collision impact, 25-30\%.
This means that the most important oxygen VUV radiation mechanism is a three-step process that consist of i) electron impact excitation to $\mathrm{O(^3P)}$ state, ii) radiative decay to $\mathrm{O(^3S)}$, iii) radiative decay to ground state and photon emission at 130\,nm.

In fact, the distribution of $I_{VUV}$ in the ($p_T$, $P_{in}$) parameter space in figure \ref{fig:VUV_power_press}, is determined by the density of O$(^3P)$.
The reason for a peak in $I_{VUV}$ is that electron impact excitation from ground state (reactions \#89-91 in table \ref{tab:electron_oxygen}) dominates the production of O$(^3P)$, and $n_e$ presents a peak in that pressure range which is consistent with the results presented in \citenum{toneli_2015}.
With increasing $p_T$ higher $n_e$ are found.
However, as the $p_T$ increases further negative ion production, mainly O$^-$, becomes more important at the expense of the electron population.
Therefore at intermediate pressures, where electron impact ionization is large and negative ion production is relatively low, the electron density finds its maximum.

%%%%%%%%%%%%%%%%%%%%%%%%%%%%%%%%%%%%%%%%%%%%%%%%%%%%%%%%%%%%%%%%%%%%%%%%%%%%%%%%
%%%%%%%%%%%%%%%%%%%%%%%%%%%%%%%%%%%%%%%%%%%%%%%%%%%%%%%%%%%%%%%%%%%%%%%%%%%%%%%%
\subsubsection{VUV emission to ion flux rate}
\begin{figure}[h]
  \centering
  \includegraphics[width=\linewidth]{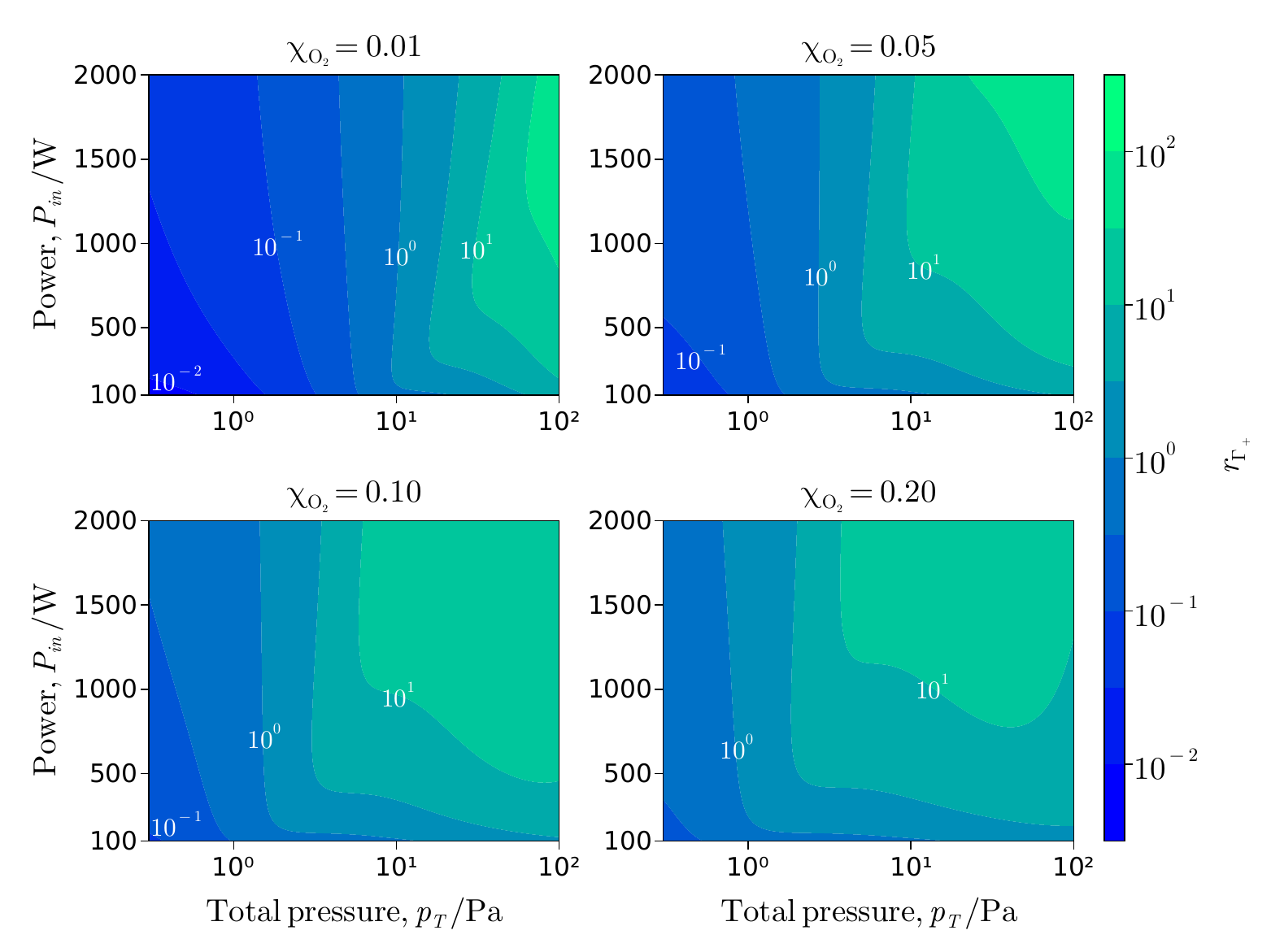}
\caption{Vacuum-ultraviolet emission intensity to positive ion flux rate, $r_{\Gamma_+}$.}
\label{fig:VUV_ion_power_press}
\end{figure}

For some industrial processes it is of interest to know photon flux, $\frac{V}{A}I_{VUV}$, with respect to the ion fluxes reaching the reactor walls, $\Gamma_+$, and therefore
\begin{equation}
r_{\Gamma_+} = \frac{\frac{A}{V}I_{VUV}}{\Gamma_+},
\end{equation}
is a useful parameter to evaluate VUV emission.
Note that $\Gamma_+ = \sum\limits_p\Gamma_p$ is the sum of the positive ion fluxes resulting from the reactions \#344-348 (in table \ref{tab:ion_flux}).
This rate is shown in figure \ref{fig:VUV_ion_power_press}.
The ion and VUV -photon fluxes are in the same order of magnitude and therefore it is possible to find operating conditions where either VUV emission dominates, $r_{\Gamma_+}\gg1$, or ion fluxes dominates, $r_{\Gamma_+}\ll1$.

\begin{figure}[h]
  \centering
  \includegraphics[width=\linewidth]{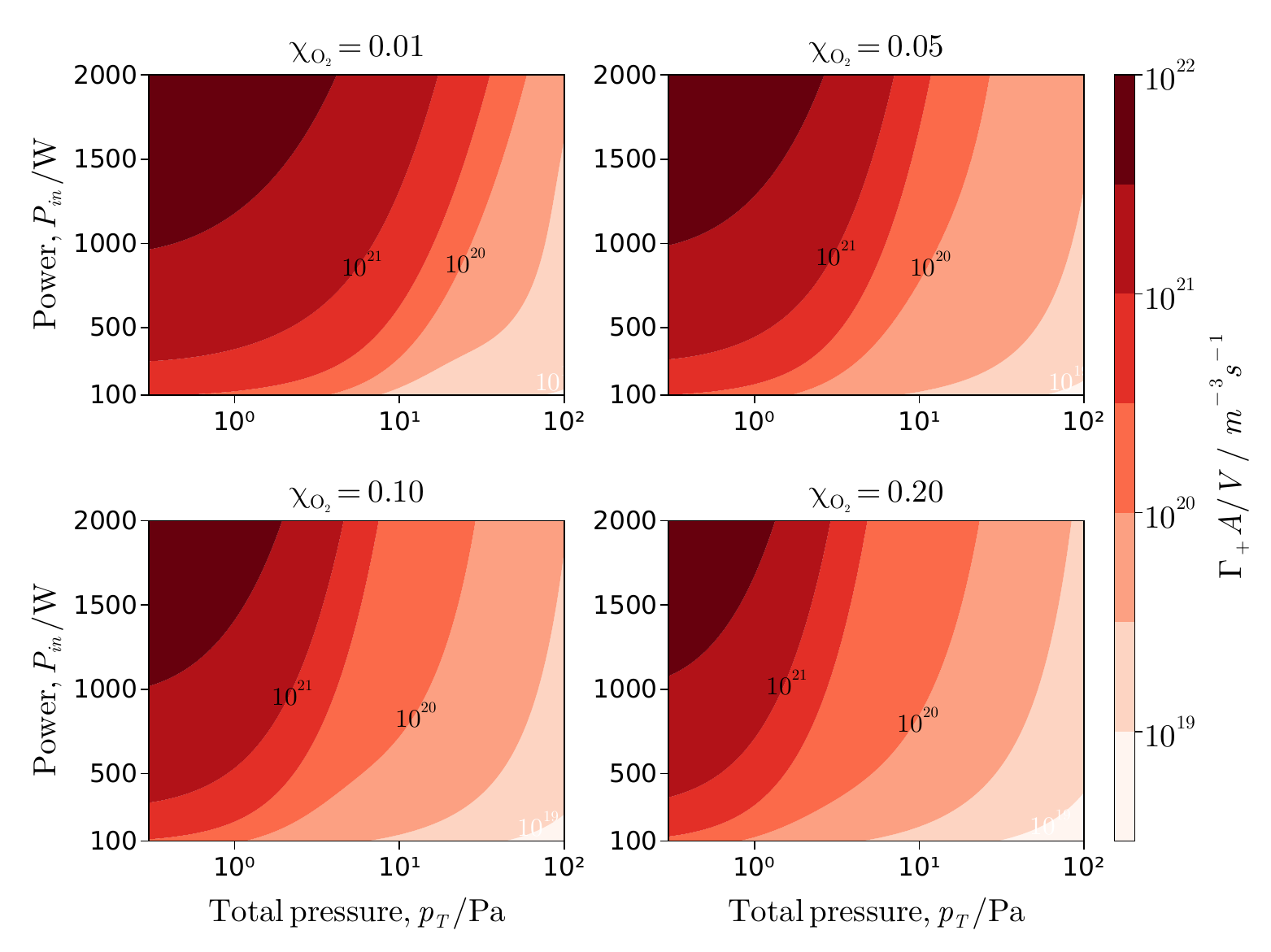}
\caption{\label{fig:ionflux_power_press}Total positive ion flux rate to the reactor walls.}
\end{figure}

The total positive ion flux, shown in figure \ref{fig:ionflux_power_press}, is strongly correlated with the plasma electronegativity $\alpha=n_-/n_e$ such that $\Gamma_+$ is largest when $\alpha\to 0$.
In general terms at lower pressures, $p_T\le 1$\,Pa, $\Gamma_+$ is large and mostly dominated by Ar$^+$, and for $p_T>10$\,Pa the electronegativity is large, $\alpha>1$, and $\Gamma_+$ drops more than an order of magnitude.
%Please refer to figures \ref{fig:Ar_ionflux} and \ref{fig:O_ionflux} in \ref{app:ion_flux_rates} for more detailed results on the fluxes of Ar$^+$, O$^+$, respectively.
This pressure dependence of $\Gamma_+$ has a significant impact on $r_{\Gamma_+}$, such that, in general terms it grows with pressure.
$r_{\Gamma_+}$ becomes largest at high $p_T$ and $P_{in}$ as in these operating conditions $I_{VUV}$ is maximum and $\Gamma_+$ drops significantly.
With increasing $\chi_{\mathrm{O}_2}$ the peak VUV intensity is displaced towards lower $p_T$, whereas $\Gamma_+$ does not change significantly, and therefore larger $r_{\Gamma_+}$ values, close to unity, are already found for $\chi_{\mathrm{O}_2}\ge 0.1$ and $p_T\sim 1$\,Pa.

%%%%%%%%%%%%%%%%%%%%%%%%%%%%%%%%%%%%%%%%%%%%%%%%%%%%%%%%%%%%%%%%%%%%%%%%%%%%%%%%
%%%%%%%%%%%%%%%%%%%%%%%%%%%%%%%%%%%%%%%%%%%%%%%%%%%%%%%%%%%%%%%%%%%%%%%%%%%%%%%%
\subsubsection{VUV emission to atomic oxygen diffusion to the wall}
The ratio between $I_{VUV}$ and atomic oxygen reaching the reactor walls may be of interest for industrial and biomedical applications as both oxygen radicals and VUV photons can readily interact with material leading to surface modifications.
This ratio is defined as follow
\begin{equation}
  r_{\mathrm{O}} = \frac{I_{VUV}}{R_{D,\mathrm{O}}},
\end{equation}
where $R_{D,\mathrm{O}} = \sum\limits_{\mathrm{O(X)}} n_{\mathrm{O(X)}}K_{D,\mathrm{O(X)}}$ is the sum of neutral diffusion reaction rates of atomic oxygen species touching the walls, i.e. reactions \#353-359 in table \ref{tab:neutral_diffusion}.

\begin{figure}[h]
  \centering
  \includegraphics[width=\linewidth]{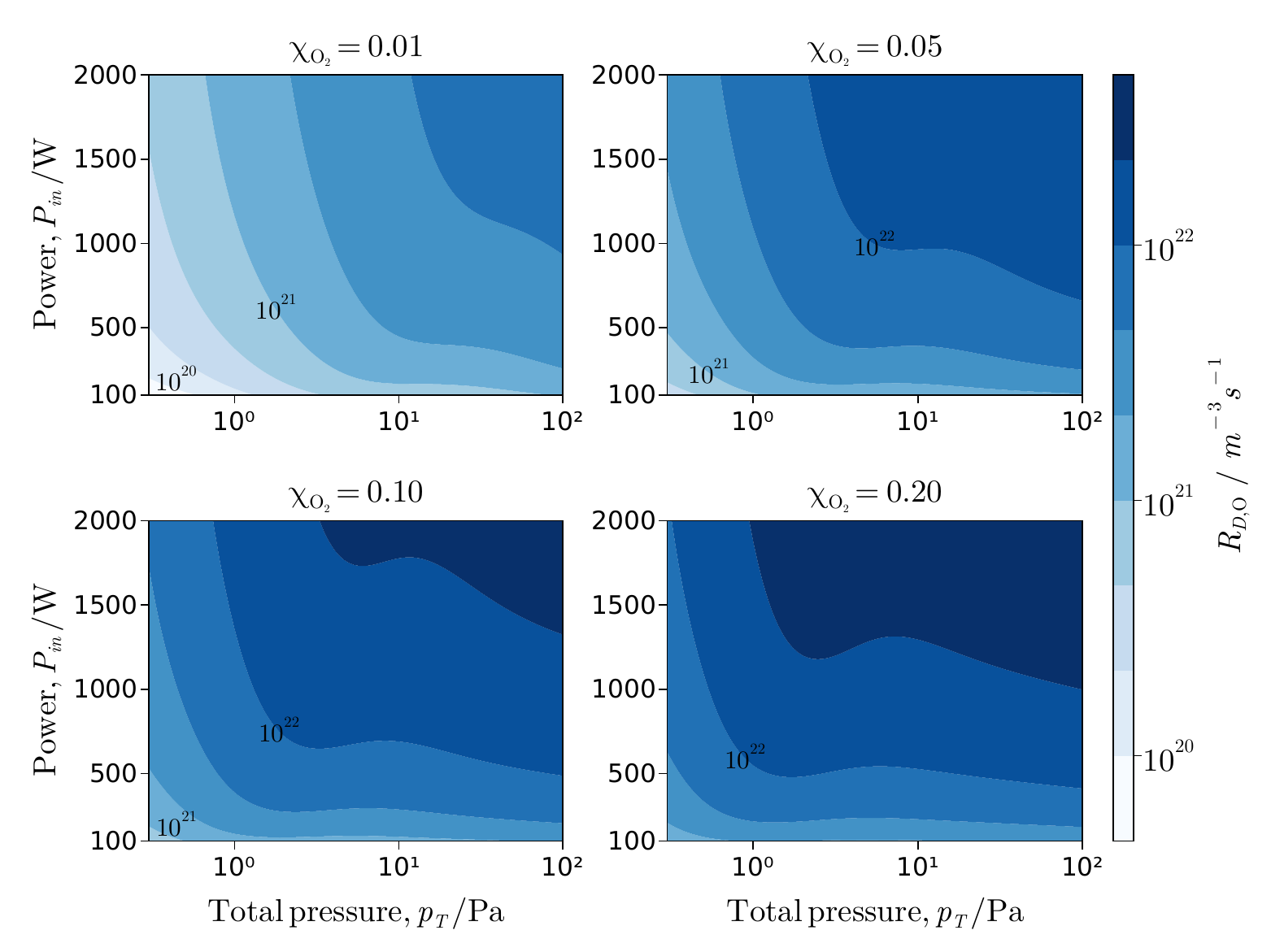}
\caption{\label{fig:Oflux_power_press}Atomic oxygen diffusion rate to the reactor walls.}
\end{figure}

First, $R_{D,\mathrm{O}}$ results are shown in figure \ref{fig:Oflux_power_press}.
The flux of oxygen radicals to the wall due to diffusion is large, especially at $p_T>10$\,Pa and $P_{in}>1000$\,W and with increasing $\chi_{\mathrm{O}_2}$.
Only at very low pressure, $<0.6$\,Pa, these fluxes can be considered low.
These trends correlate mainly with atomic oxygen density, which presents a similar distribution in the parameter space investigated.

\begin{figure}[h]
  \centering
  \includegraphics[width=\linewidth]{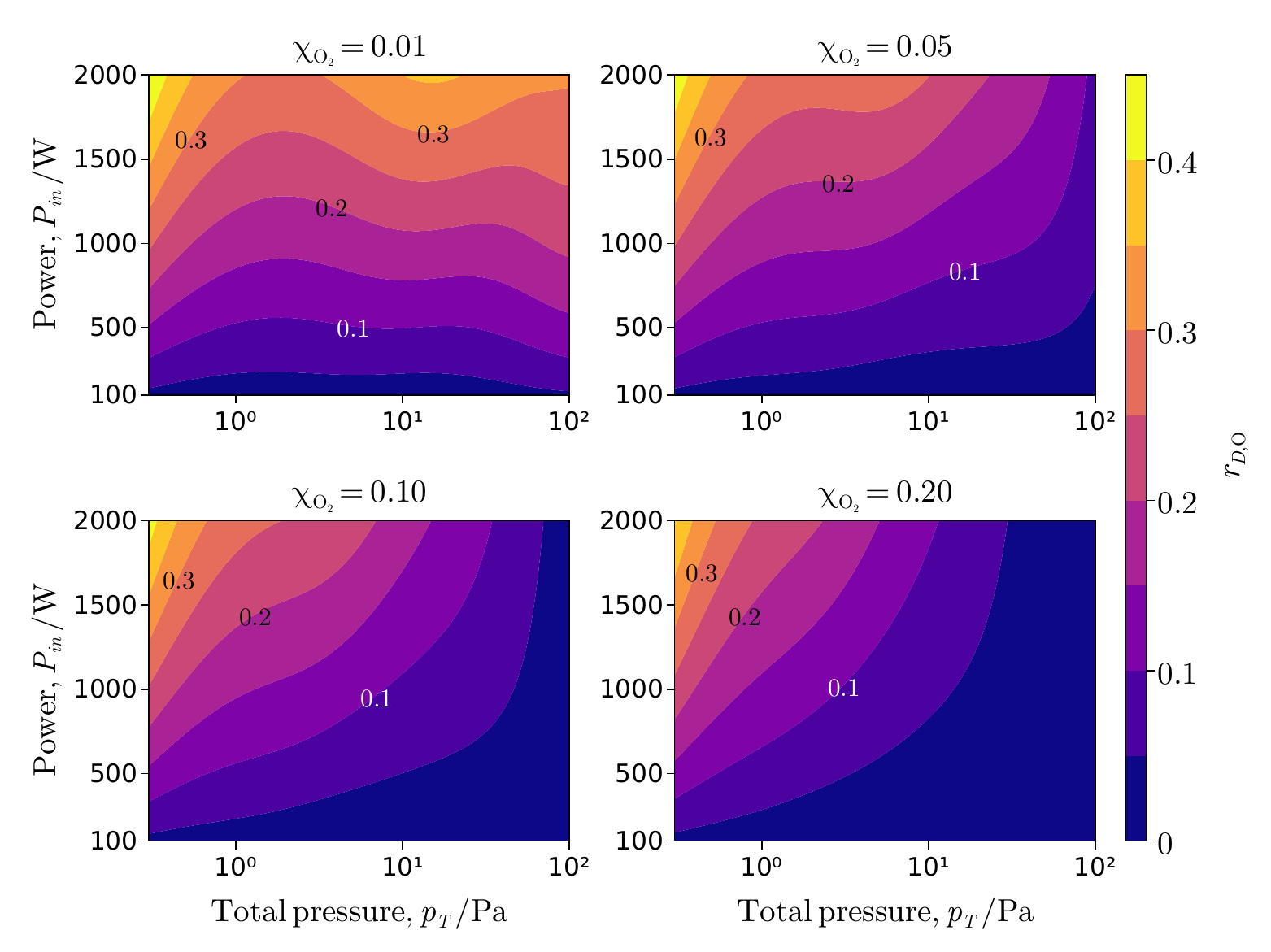}
\caption{Vacuum-ultraviolet emission to atomic oxygen surface flux rate.}
\label{fig:VUV_oxygen_power_press}
\end{figure}

The results for $r_\mathrm{O}$ are presented in figure \ref{fig:VUV_oxygen_power_press}.
This data shows that $I_{VUV}$ is always lower than $R_{D,\mathrm{O}}$.
The maximum values, $r_\mathrm{O}\sim0.4$, are found at minimum $p_T\sim$0.3\,Pa, and maximum power,$P_{in}\sim2000$\,W and decreases with increasing $\chi_{\mathrm{O}_2}$.
The minimum values, $r_\mathrm{O}\to0$, are found in a larger region of high $p_T$ and low $p_T$.

%%%%%%%%%%%%%%%%%%%%%%%%%%%%%%%%%%%%%%%%%%%%%%%%%%%%%%%%%%%%%%%%%%%%%%%%%%%%%%%%
%%%%%%%%%%%%%%%%%%%%%%%%%%%%%%%%%%%%%%%%%%%%%%%%%%%%%%%%%%%%%%%%%%%%%%%%%%%%%%%%
%%%%%%%%%%%%%%%%%%%%%%%%%%%%%%%%%%%%%%%%%%%%%%%%%%%%%%%%%%%%%%%%%%%%%%%%%%%%%%%%
\section{Summary}
In this work we have conducted a numerical investigation of oxygen VUV emission in Ar/O$_2$ DICP.
For this purpose we have developed a 0D plasma chemical-kinetics GM that implements an extended chemical-radiative reaction scheme for Ar and O$_2$ species.
The first part of the results investigates Ar/O$_2$ DICP for operating parameters between 200-800\,W, 2-20\,Pa and 0-0.20 O$_2$ fractions.
Moreover, because the GM works with a fixed temperature $T_N$ for neutrals and ions, $T_N$ has also been varied between 400 and 2000 K to test the impact of $T_N$ on the plasma results.
The numerical results have been presented alongside experimental work conducted specifically for this investigation.
The results show that the GM is performing correctly and that $T_N$ does have an impact on the final results but within a relatively small range.
The gas and plasma results, as well as the emission lines measured are as expected although some differences are observed for argon metastables.
The source of these discrepancies is not yet clear, as they are not necessarily errors in the numerical method, and thus results are taken as valid.
Oxygen VUV emission results show good agreement, with the 130\,nm line, from the $\mathrm{O(^3S)}\to \mathrm{O}$ transition, clearly dominating.
The 135\,nm line, from the $\mathrm{O(^5S)}\to \mathrm{O}$ transition, is an order of magnitude lower and emission from cascading reactions is negligible.

The second part of results investigates oxygen VUV emission over a broader range of total pressure and power.
The GM results for plasmas with 0.3-100\,Pa and 100-2000\,W have shown that oxygen VUV emission, in general terms, increases within the investigated power and oxygen fraction and peak emission intensities are found for pressures between 5-50\,Pa.
The 130\,nm line dominates for most of the parameter space investigated.
Surprisingly the most frequent chemical pathway that generates O$(^3S)$ is not direct electron impact excitation from ground state, but excitation to O$(^3P)$ that then decays to O$(^3S)$.

Results of VUV emission intensities with respect to ion fluxes and oxygen diffusion to the reactor walls have also been presented.
While VUV emission is largest with respect to ion fluxes at high pressures, oxygen diffusion is much larger than VUV emission for the parameter space investigated.

\section{Acknowledgments}
This project was undertaken on the Viking Cluster, which is a high performance computing facility provided by the University of York. We are grateful for computational support from the University of York High Performance Computing service, Viking and the Research Computing team.
The authors wish to acknowledge financial support from the EPSRC Centre for Doctoral Training in Fusion Energy (EP/L01663X/1) and the UKRI Engineering and Physical Sciences. This has also been funded by the Deutsche Forschungsgemeinschaft (DFG, German
Research Foundation) – project number 424927143.
We also thank the team at House of Plasma GmbH for support in using the Multipole Resonance Probe.

% Comment on pressure limitations of the model
\clearpage
\section*{References}
\bibliography{paper}

\clearpage
\appendix

%%%%%%%%%%%%%%%%%%%%%%%%%%%%%%%%%%%%%%%%%%%%%%%%%%%%%%%%%%%%%%%%%%%%%%%%%%%%%%%%%%%%%
%%%%%%%%%%%%%%%%%%%%%%%%%%%%%%%%%%%%%%%%%%%%%%%%%%%%%%%%%%%%%%%%%%%%%%%%%%%%%%%%%%%%%
\section{Plasma-chemical reaction scheme}
\label{app:reaction_scheme}
Please note that the rate coefficients for the reactions from \citenum{fiebrandt_2020, fiebrandt_thesis}  were generated assuming a Maxwellian energy distribution function (EDF) for electrons with temperatures between 1.5 and 4\,eV.
%%%%%%%%%%%%%%%%%%%%%
%% ELECTRON OXYGEN %%
%%%%%%%%%%%%%%%%%%%%%
\scriptsize
\begin{longtable}[c]{@{}lllll}
 \caption{\label{tab:electron_oxygen}Electron-oxygen reactions. Electron temperature, $T_e$, in eV and neutral temperature, $T_N$, in K. $N_r$ is the number of reactants.}\\
 		\br
 		\textbf{\#} & \textbf{Reaction} & \textbf{E$_{thr}$ [eV]} & \textbf{$K_r$ [m$^{3+3(N_r-2)}$s$^{-1}$]} & \textbf{Ref.} \\
 		\mr
 \endfirsthead

 \multicolumn{5}{c}{Continuation of table \ref{tab:electron_oxygen}: Electron-oxygen reactions.}\\
 		\br
 		\textbf{\#} & \textbf{Reaction} & \textbf{E$_{thr}$ [eV]} & \textbf{$K_r$ [m$^{3+3(N_r-2)}$s$^{-1}$]} & \textbf{Ref.} \\
 		\mr
 \endhead

 		\br
 \endfoot
 		\br
 \endlastfoot

 %Data
      1 & $e + \mathrm{O} \to 2e + \mathrm{O}^+ $ & 13.6 & $4.93\cdot 10^{-15} T_e^{ 0.723} \exp(-13.20/T_e) $ & \cite[reaction 12]{turner_2015}, \citenum{laher_1990}\cr
2 & $e + \mathrm{O} \to e + \mathrm{O}(^1\mathrm{D}) $ & 1.96 & $8.45\cdot 10^{-15} T_e^{-0.306} \exp( -3.13/T_e) $ & \cite[reaction 13]{turner_2015}, \citenum{laher_1990}\cr
3 & $e + \mathrm{O} \to e + \mathrm{O}(^1\mathrm{S}) $ & 4.18 & $1.04\cdot 10^{-15} T_e^{-0.134} \exp( -4.19/T_e) $ & \cite[reaction 14]{turner_2015}, \citenum{laher_1990}\cr
4 & $e + \mathrm{O}(^1\mathrm{D}) \to 2e + \mathrm{O}^+ $ & 11.65 & $4.93\cdot 10^{-15} T_e^{ 0.723} \exp(-11.64/T_e) $ & \cite[reaction 15]{turner_2015}\cr
5 & $e + \mathrm{O}(^1\mathrm{D}) \to e + \mathrm{O} $ & -1.96 & $8.45\cdot 10^{-15} T_e^{ 0.306} \exp( -1.17/T_e) $ & \cite[reaction 16]{turner_2015}, \citenum{laher_1990}\cr
6 & $e + \mathrm{O}(^1\mathrm{S}) \to 2e + \mathrm{O}^+ $ & 9.43 & $4.93\cdot 10^{-15} T_e^{ 0.723} \exp( -9.42/T_e) $ & \cite[reaction 17]{turner_2015}\cr
7 & $e + \mathrm{O}(^1\mathrm{S}) \to e + \mathrm{O} $ & -4.18 & $1.04\cdot 10^{-15} T_e^{-0.134} \exp( -0.73/T_e) $ & \cite[reaction 18]{turner_2015}, \citenum{laher_1990}\cr
8 & $e + \mathrm{O}^- \to 2e + \mathrm{O} $ & 3.44 & $9.33\cdot 10^{-14} T_e^{ 0.178} \exp( -3.13/T_e) $ & \cite[reaction 19]{turner_2015}, \citenum{deutsch_2003}\cr
\end{longtable}
\normalsize

%%%%%%%%%%%%%%%%%%%%
%% ELECTRON ARGON %%
%%%%%%%%%%%%%%%%%%%%
\clearpage
\begin{table}[h]
  \caption{\label{tab:electron_argon}Electron-argon reactions. Electron temperature, $T_e$, in eV, and $N_r$ is the number of reactants.}
  \centering
	\begin{tabular}{@{}lllll}
 		\br
 		\textbf{\#} & \textbf{Process} & \textbf{E$_{thr}$ [eV]} & \textbf{$K_r$ [m$^{3+3(N_r-2)}$s$^{-1}$]} & \textbf{Ref.} \\
 		\mr
      %  # Ar reactions (Gudmundsson 2007, table 1)
107 & $e + \mathrm{Ar} \to e + \mathrm{Ar} $ & 0.0 & $2.336\cdot 10^{-14} T_e^{1.609}\cdot$ & \citenum{hurlbatt_2017,gudmundsson_2002} \cr
     &                                        &        & $\quad\exp{\big[0.0618  (\log{T_e})^2 - 0.1171(\log{T_e})^3 \big]}$ &  \cr
108 & $e + \mathrm{Ar} \to \mathrm{Ar}^+ + 2e $ & 15.76 & $ 2.3\cdot 10^{-14} T_e^{0.59} \exp(-17.44/T_e) $ & \citenum{gudmundsson_2007,straub_1995}\cr
109 & $e + \mathrm{Ar} \to \mathrm{Ar}^m + e $ & 11.55 & $ 5.0\cdot 10^{-15} \exp(-12.64 / T_e) $ & \citenum{gudmundsson_2007,tachibana_1986}\cr
110 & $e + \mathrm{Ar} \to \mathrm{Ar}^m + e $ & 11.72 & $ 1.4\cdot 10^{-15} \exp(-12.42 / T_e) $ & \citenum{gudmundsson_2007,tachibana_1986}\cr
111 & $e + \mathrm{Ar} \to \mathrm{Ar}^r + e $ & 11.62 & $ 1.9\cdot 10^{-15} \exp(-12.60 / T_e) $ & \citenum{gudmundsson_2007,tachibana_1986}\cr
112 & $e + \mathrm{Ar} \to \mathrm{Ar}^r + e $ & 11.83 & $ 2.7\cdot 10^{-16} \exp(-12.14 / T_e) $ & \citenum{gudmundsson_2007,tachibana_1986}\cr
\end{tabular}
\end{table}

%%%%%%%%%%%%%%%%%%%
%% OXYGEN OXYGEN %%
%%%%%%%%%%%%%%%%%%%
\clearpage
\scriptsize
\begin{longtable}[c]{@{}llll}
 \caption{\label{tab:oxygen_oxygen}Oxygen-oxygen reactions. Electron temperature, $T_e$, in eV and neutral and ion temperature, $T_N$, in K. $N_r$ is the number of reactants.}\\
 		\br
 		\textbf{\#} & \textbf{Reaction} & \textbf{$K_r$ [m$^{3+3(N_r-2)}$s$^{-1}$]} & \textbf{Ref.} \\
 		\mr
 \endfirsthead

 \multicolumn{4}{c}{Continuation of table \ref{tab:oxygen_oxygen}: Oxygen-oxygen reactions}\\
 		\br
 		\textbf{\#} & \textbf{Reaction} & \textbf{$K_r$ [m$^{3+3(N_r-2)}$s$^{-1}$]} & \textbf{Ref.} \\
 		\mr
 \endhead

 		\br
 \endfoot

    \br
    \multicolumn{4}{ l }{\multirow{1}{=}{$^a$ The collisional quenching coefficient for $\mathrm{O(^1D)} + \mathrm{O_2}$ is measured in Ref. \citenum{demore_1997}. Here, the same quenching coefficient is used for the marked reactions, due to a lack of specific data.}}\cr\cr
    \multicolumn{4}{ l }{\multirow{1}{=}{$^b$ The collisional quenching coefficient for $\mathrm{O(^1S)} + \mathrm{O_2}$ is measured in Refs. \citenum{bortner_1978,slanger_1978}. Here, the same quenching coefficient is used for the marked reactions, due to a lack of specific data.}}\cr\cr
    \multicolumn{4}{ l }{\multirow{2}{=}{$^c$ The collisional quenching coefficient for $\mathrm{O(^3S)} + \mathrm{O_2}$ is measured in Ref. \citenum{niemi_2005}. Here, the same quenching coefficient is used for the marked reactions, due to a lack of specific data.}}\cr\cr
    \multicolumn{4}{ l }{\multirow{2}{=}{$^d$ The collisional quenching coefficient for $\mathrm{O(^5P)} + \mathrm{O_2}$ is measured in Ref. \citenum{dagdigian_1988}. Here, the same quenching coefficient is used for the marked reactions, due to a lack of specific data.}}\cr\cr
    \multicolumn{4}{ l }{\multirow{2}{=}{$^e$ The collisional quenching coefficient for $\mathrm{O(^5S)} + \mathrm{O_2}$ is measured in Ref. \citenum{roth_1973}. Here, the same quenching coefficient is used for the marked reactions, due to a lack of specific data.}}
 \endlastfoot

 %Data
      125 & $3\mathrm{O} \to \mathrm{O} + \mathrm{O}_2 $ & $3.80\cdot 10^{-44} (300/T_N) \exp(-170/ T_N) $ & \cite[reaction 92]{turner_2015}, \citenum{johnston_1968}\cr
126 & $3\mathrm{O} \to \mathrm{O} + \mathrm{O}_2(\mathrm{b^1}\Sigma_u^+) $ & $1.40\cdot 10^{-42} \exp(-650/T_N) $ & \cite[reaction 93]{turner_2015}, \citenum{bortner_1978}\cr
127 & $2\mathrm{O} + \mathrm{O}_2 \to \mathrm{O} + \mathrm{O}_3 $ & $4.20\cdot 10^{-47} \exp(1056/T_N) $ & \cite[reaction 94]{turner_2015}, \citenum{johnston_1968,rawlins_1987}\cr
128 & $2\mathrm{O} + \mathrm{O}_2 \to \mathrm{O} + \mathrm{O}_3(\nu) $ & $9.80\cdot 10^{-47} \exp(1056/T_N) $ & \cite[reaction 95]{turner_2015}, \citenum{johnston_1968,rawlins_1987}\cr
129 & $2\mathrm{O} + \mathrm{O}_2 \to \mathrm{O}_2(\mathrm{a^1}\Delta_u) + \mathrm{O}_2 $ & $6.50\cdot 10^{-45} (300/T_N) \exp(-170/T_N) $ & \cite[reaction 96]{turner_2015}, \citenum{johnston_1968,slanger_2003}\cr
130 & $2\mathrm{O} + \mathrm{O}_2 \to \mathrm{O}_2(\mathrm{b^1}\Sigma_u^+) + \mathrm{O}_2 $ & $6.50\cdot 10^{-45} (300/T_N) \exp(-170/T_N) $ & \cite[reaction 97]{turner_2015}, \citenum{johnston_1968,slanger_2003}\cr
131 & $\mathrm{O} + \mathrm{O}(^1\mathrm{D}) \to 2\mathrm{O} $ & $2.00\cdot 10^{-18} $ & \cite[reaction 98]{turner_2015}, \citenum{sobral_1993}\cr
132 & $\mathrm{O} + \mathrm{O}(^1\mathrm{S}) \to 2\mathrm{O} $ & $2.50\cdot 10^{-17} \exp(-300/T_N) $ & \cite[reaction 99]{turner_2015}, \citenum{schofield_1978}\cr
\end{longtable}
\normalsize

%%%%%%%%%%%%%%%%%
%% ARGON ARGON %%
%%%%%%%%%%%%%%%%%
\clearpage
\begin{table}[h]
  \caption{\label{tab:argon_argon}Argon-argon reactions. Electron temperature, $T_e$, in eV and neutral and ion temperature, $T_N$, in K. $N_r$ is the number of reactants.}
  \centering
	\begin{tabular}{@{}llll}
 		\br
 		\textbf{\#} & \textbf{Reaction} & \textbf{$K_r$ [m$^{3+3(N_r-2)}$s$^{-1}$]} & \textbf{Ref.} \\
 		\mr
      % ARGON-ARGON
259 & $2\mathrm{Ar}^m \to 2\mathrm{Ar} $ & $ 2.0\cdot 10^{-13} $ & \citenum{gudmundsson_2007}\cr
260 & $\mathrm{Ar}^m + \mathrm{Ar}^r \to \mathrm{Ar} + \mathrm{Ar}^+ + e $ & $ 2.1\cdot 10^{-15} $ & \citenum{gudmundsson_2007,bassett_1994}\cr
261 & $\mathrm{Ar(4p)} + \mathrm{Ar(4p)} \to \mathrm{Ar} + \mathrm{Ar}^+ + e$ & $ 5.0\cdot 10^{-16} $ & \citenum{gudmundsson_2007,kannari_1985}\cr
262 & $2\mathrm{Ar}^m \to \mathrm{Ar} + \mathrm{Ar}^+ + e $ & $ 6.4\cdot 10^{-16} $ & \citenum{gudmundsson_2007,ferreira_1985}\cr
\end{tabular}
\end{table}

%%%%%%%%%%%%%%%%%%
%% OXYGEN ARGON %%
%%%%%%%%%%%%%%%%%%
\clearpage
\begin{longtable}[c]{@{}llll}
 \caption{\label{tab:argon_oxygen}Argon-oxygen reactions. Electron temperature, $T_e$, in eV and neutral and ion temperature, $T_N$, in K. $N_r$ is the number of reactants.}\\
 		\br
 		\textbf{\#} & \textbf{Reaction} & \textbf{$K_r$ [m$^{3+3(N_r-2)}$s$^{-1}$]} & \textbf{Ref.} \\
 		\mr
 \endfirsthead

 \multicolumn{4}{c}{Continuation of table \ref{tab:argon_oxygen}: Argon-oxygen reactions.}\\
 		\br
 		\textbf{\#} & \textbf{Reaction} & \textbf{$K_r$ [m$^{3+3(N_r-2)}$s$^{-1}$]} & \textbf{Ref.} \\
 		\mr
 \endhead

 		\br
 \endfoot
 		\br
    \multicolumn{4}{ l }{\footnotesize$^a$ The reaction is an extension from the reactions in Ref. \citenum{gudmundsson_2007}.} \cr
    \multicolumn{4}{ l }{\multirow{2}{=}{\footnotesize$^b$ The collisional quenching coefficient for $\mathrm{O(^3P)} + \mathrm{Ar}$ is measured in Ref. \citenum{niemi_2005}. Here, the same quenching coefficient is used for the marked reactions, due to a lack of specific data.}}
 \endlastfoot

 %Data
      %  # Ar/O_2 reactions. Table 7 (Gudmundsson 2007)                                                                                                             
264 & $\mathrm{O} + \mathrm{Ar}^m \to \mathrm{O} + \mathrm{Ar} $ & $ 4.1 \cdot 10^{-17} $ & \citenum{gudmundsson_2007,king_1977}\cr
265 & $\mathrm{O} + \mathrm{Ar}^r \to \mathrm{O} + \mathrm{Ar} $ & $ 4.1 \cdot 10^{-17} $ & \citenum{gudmundsson_2007}$^a$\cr
266 & $\mathrm{O}_2 + \mathrm{Ar(4p)} \to \mathrm{O} + \mathrm{O} + \mathrm{Ar} $ & $ 2.96\cdot 10^{-16} $ & \citenum{gudmundsson_2007}\cr
267 & $\mathrm{O}_2(\mathrm{a^1}\Delta_u) + \mathrm{Ar(4p)} \to \mathrm{O} + \mathrm{O} + \mathrm{Ar} $ & $ 2.96\cdot 10^{-16} $ &\citenum{gudmundsson_2007}$^a$ \cr
268 & $\mathrm{O}_2(\mathrm{b^1}\Sigma_u^+) + \mathrm{Ar(4p)} \to \mathrm{O} + \mathrm{O} + \mathrm{Ar} $ & $ 2.96\cdot 10^{-16} $ &\citenum{gudmundsson_2007}$^a$ \cr
269 & $\mathrm{O}_2 + \mathrm{Ar(4p)} \to \mathrm{O} + \mathrm{O}\mathrm{(^1D)} + \mathrm{Ar} $ & $ 3.34\cdot 10^{-16} $ & \citenum{gudmundsson_2007}\cr
270 & $\mathrm{O}_2(\mathrm{a^1}\Delta_u) + \mathrm{Ar(4p)} \to \mathrm{O} + \mathrm{O}\mathrm{(^1D)} + \mathrm{Ar} $ & $ 3.34\cdot 10^{-16} $ &\citenum{gudmundsson_2007}$^a$ \cr
271 & $\mathrm{O}_2(\mathrm{b^1}\Sigma_u^+) + \mathrm{Ar(4p)} \to \mathrm{O} + \mathrm{O}\mathrm{(^1D)} + \mathrm{Ar} $ & $ 3.34\cdot 10^{-16} $ & \citenum{gudmundsson_2007}$^a$\cr
\end{longtable}

%%%%%%%%%%%%%%%%%%%
%% RECOMBINATION %%
%%%%%%%%%%%%%%%%%%%
\clearpage
\begin{longtable}[c]{@{}llll}
  \caption{\label{tab:recombination} Recombination reactions. Neutral and ion temperature, $T_N$, in K, and $N_r$ is the number of reactants.}\\
 		\br
 		\textbf{\#} & \textbf{Reaction} & \textbf{$K_r$ [m$^{3+3(N_r-2)}$s$^{-1}$]} & \textbf{Ref.} \\
 		\mr
 \endfirsthead

 \multicolumn{4}{c}{Continuation of table \ref{tab:recombination}: Recombination reactions.}\\
 		\br
 		\textbf{\#} & \textbf{Reaction} & \textbf{$K_r$ [m$^{3+3(N_r-2)}$s$^{-1}$]} & \textbf{Ref.} \\
 		\mr
 \endhead

 		\br
 \endfoot

  %\multicolumn{2}{| c |}{End of Table}\\
 		\br
 \endlastfoot

 %Data
      % Recombination
302 & $\mathrm{O}^+ + \mathrm{O}^- \to 2\mathrm{O} $ & $3.10\cdot 10^{-14} (300/T_N)^{1.1} $ & \cite[reaction 186]{turner_2015}, \citenum{miller_2012}\cr
303 & $\mathrm{O}^+ + \mathrm{O}^- + \mathrm{O}_2 \to 2\mathrm{O} + \mathrm{O}_2 $ & $1.00\cdot 10^{-37} (300/T_N)^{2.5} $ & \cite[reaction 187]{turner_2015}, \citenum{biondi_1969}\cr
304 & $\mathrm{O}^+ + \mathrm{O}^- + \mathrm{O}_2 \to 2\mathrm{O}_2 $ & $1.00\cdot 10^{-37} (300/T_N)^{2.5} $ & \cite[reaction 188]{turner_2015}, \citenum{biondi_1969}\cr
305 & $\mathrm{O}^- + \mathrm{O}_2^+ \to 3\mathrm{O} $ & $1.61\cdot 10^{-14} (300/T_N)^{1.1} $ & \cite[reaction 207]{turner_2015}, \citenum{miller_2012}\cr
306 & $\mathrm{O}^- + \mathrm{O}_2^+ \to \mathrm{O} + \mathrm{O}_2 $ & $1.61\cdot 10^{-14} (300/T_N)^{1.1} $ & \cite[reaction 208]{turner_2015}, \citenum{miller_2012}\cr
307 & $\mathrm{O}^- + \mathrm{O}_3^+ \to \mathrm{O} + \mathrm{O}_3 $ & $3.07\cdot 10^{-14} (300/T_N)^{1.1} $ & \cite[reaction 212]{turner_2015}, \citenum{miller_2012}\cr
308 & $\mathrm{O}^- + \mathrm{O}_4^+ \to \mathrm{O} + 2\mathrm{O}_2 $ & $1.54\cdot 10^{-14} (300/T_N)^{0.9} $ & \cite[reaction 213]{turner_2015}, \citenum{miller_2012}\cr
309 & $\mathrm{O}^- + \mathrm{O}_4^+ \to \mathrm{O}_2 + \mathrm{O}_3 $ & $1.54\cdot 10^{-14} (300/T_N)^{0.9 } $ & \cite[reaction 214]{turner_2015}, \citenum{miller_2012}\cr
\end{longtable}

\end{document}